\documentclass[a4paper,10pt]{elsarticle}

\usepackage{graphicx}
\usepackage{amsfonts}
\usepackage{amssymb}
\usepackage{fullpage}
\usepackage{url}
        {\begin{list}{ {\bf Proof}}%
                           {\setlength{\leftmargin}{\labelwidth}} \item }%
        {\end{list}}

\newtheorem{definition}{Definition}

\begin{document}

\begin{frontmatter}

\title{\sc Busy Beaver Machines and the Observant Otter Heuristic \\ (or How to Tame Dreadful Dragons)}
\author{James Harland}
\address{School of Computer Science and Information Technology \\ 
RMIT University, GPO Box 2476 \\ 
Melbourne, 3001, Australia \\ 
{\em james.harland@rmit.edu.au} }

\begin{abstract}
The busy beaver is a well-known specific example of a non-computable function. Whilst many aspect of this problem have been investigated, it is not always easy to find thorough and convincing evidence for the claims made about the maximality of particular machines, and the phenomenal size of some of the numbers involved means that it is not obvious that the problem can be feasibly addressed at all. In this paper we address both of these issues. We discuss a framework in which the busy beaver problem and similar problems may be addressed, and the appropriate processes for providing evidence of claims made. We also show how a simple heuristic, which we call the \textit{observant otter}, can be used to evaluate machines with an extremely large number of execution steps required to terminate. We also show empirical results for an implementation of this heuristic which show how this heuristic is effective for all known `monster' machines. 
\end{abstract}


\end{frontmatter}

\section{Introduction}

The \textit{busy beaver problem} has been an object of fascination since its introduction by Rado in 1962 as a specific example of a non-computable function \cite{Rado62}. The problem is to find the largest number of non-blank characters that are printed by a terminating Turing machine of no more than a given size on the blank input. Given the simplicity of the problem statement, it is often surprising to discover the extraordinarily large numbers of symbols that can be printed by some machines despite containing only a handful of states
(see Table~\ref{table1} below; the number of non-blank characters printed by a machine is known as its \textit{productivity} \cite{BBJ}). 
It is counterintuitive, to say the least, to learn that a machine with only six states may terminate on the blank input after $10^{36,534}$ steps! The sheer size of this and similar numbers is not only motivation for an analysis of their behaviour, but also means that naive execution of such machines is hopelessly infeasible. 

\begin{table}
\begin{center}
 \begin{tabular}{cccll}
\textbf{States} & \textbf{Symbols} & \textbf{States} $\times$ \textbf{Symbols} & \textbf{Non-blank characters} & \textbf{Hops} \\
1 & 2 & 2 & 1 & 1 \\
2 & 2 & 4 & 4 & 6 \\
3 & 2 & 6 & 6 & 21 \\
2 & 3 & 6 & 9 & 38  \\
2 & 4 & 8 & $\geq$ 2050 & $\geq$ 3,932,964 \\
4 & 2 & 8 & 13 & 107 \\
3 & 3 & 9 & $\geq$ 374,676,383 & $\geq 1.12 * 10^{18}$ \\
2 & 5 & 10 & $\geq 1.7*10^{352}$ & $\geq 1.9*10^{704}$ \\
5 & 2 & 10 & $\geq 4098$ & $\geq 47,176,870$ \\
2 & 6 & 12 & $\geq 1.9*10^{4933}$ & $\geq 2.4*0^{9866}$ \\
3 & 4 & 12 & $\geq 3.7*0^{6518}$ & $\geq 5.2*10^{13036}$ \\
4 & 3 & 12 & $\geq 1.383*10^{7036}$ & $\geq 1.025*10^{14072}$ \\
6 & 2 & 12 & $\geq 3.51*10^{18267}$ & $\geq 7.41*10^{36534}$ \\
\end{tabular}
\caption{``Dreadful dragon'' records for various classes of machines}
\label{table1}
\end{center}
\end{table}

Whilst analyses of specific machines can reveal surprising and interesting properties \cite{LP07,Michel}, in order to determine the maximum value for a given size of machine, it is necessary to search through all such machines and record the maximum value found. In order to make such a result scientifically credible, this should not be simply an algorithmic search which produces a conclusion, but also sufficient evidence to allow the result to be checked or reproduced. Unfortunately, this is generally not the case for the known results for the busy beaver. For example, Lin and Rado \cite{LR65} analyse the case for 3 states and 2 symbols by using a program to reduce the unknown cases (or `holdouts') to 40, and ``.. these 40 holdouts were checked by hand''. Whilst they provide a specification of the 40 holdout machines and an illustration of their method, the details provided are descriptive rather than comprehensive. Similarly Brady describes the determination of busy beaver function for the case of 4 states and 2 symbols by a process of search through a total of around 500,000 machines, from which there remained a total of 5,280 remained unclassified. Some specially designed programs further reduced the holdouts to a list of 218. Brady then states ``Direct inspection of each of the 218 holdouts from these programs reveals that none will ever halt, ...'' (\cite{Brady83}, p.649), and uses this and other results to determine the busy beaver value. In other words, the remaining 218 cases were hand checked, but without any evidence provided. Machlin and Stout, who performed a similar analysis of the same class of machines, also use programs to significantly aid their search, until 210 holdouts remain \cite{MS90}. However, similar comments apply to their statement  that ``The final 210 holdouts were examined by hand to verify that they were in infinite loops.'' (\cite{MS90}, p.95). There also appears to be no record of the code or data files used by Lin and Rado, Brady or Machlin and Stout in these searches. 

Perhaps the most comprehensive analysis of the busy beaver problem to date is that by Lafitte and Papazian \cite{LP07}. They have analysed various instance of the busy beaver problem, including an enumeration of all machines with 2 states and 3 symbols, and those with 3 states and 2 symbols. They have also provided analyses of some of the larger cases (2 states 4 symbols, 4 states 2 symbols, 3 states 3 symbols), but in less detail, and have identified some problematic machines. This is level of detail is a good start, but unfortunately falls short in terms of providing reproducable evidence, and as above, there appears to be no code available, nor any data files. 

Similar remarks apply to the work of Wolfram \cite{Wolfram84,Wolfram}. His work is wider in scope, in that it takes in various properties of a number of types of automata, but the process he follows is comparable, in that it is based on searching through large numbers of machines looking for properties of interest. However, the ability to reproduce (and perhaps vary slightly) the process that has been undertaken seems difficult based on the information provided, especially as the busy beaver problem for Turing machines forms a small part of his overall work. 

Whilst there is no reason whatsoever to doubt the veracity or sincerity of these results and others like them in the literature, it seems unscientific to accept such results as proven in the absence of both mathematical proof and empirical evidence that can be inspected, assessed and checked. This seems particularly true for claims about the maximality of specific machines, and hence particular values of the busy beaver function. In fairness, it should be said that much of the above work was done when computing resources were significantly more limited than today, and over the course of more than seventy years now there have been phenomenal improvements in processing power, storage capacity and network bandwidth which have made tasks that were once unthinkable into ones which are now routine. Those same improvements mean that now, in the era of cloud computing (in which data and computation can be easily distributed), it seems that the provision of such `computational evidence' should be a minimum requirement for such claims. 
\textbf{This evidence should include not only the programs used for the search, but also the list of machines generated as well as the evidence used to draw conclusions about their status.} 
In other words, it is not sufficient just to determine the appropriate values; we must do so in such a way that provides verifiable and reproducable evidence for why this is the case. This seems particularly important now, when the famous proof of the Four Colour Theorem seems like ancient history \cite{AH77}, and the much more recent work of Hales et al.\ on the Flyspeck project to prove the Kepler conjecture represents the cutting edge of mathematical knowledge incorporating computation \cite{Hales05,Flyspeck}. 

The nature of the process for determining busy beaver values has been discussed by de Mol\cite{deMol}. Her aim is different to ours, in that she is interested in the cultural and philosophical aspects of the nature of computer-assisted proof in mathematics. de Mol discusses the issue of evidence in such cases, and the problem of \textit{unsurveyability} (i.e.\ that for many computer-assisted proofs, it is possible for humans to comprehend all details of the proof), and the importance of independent verification of results. From our perspective, the main conclusion is that computer-assisted proofs (which include the busy beaver problem and its variants) should include a \textit{description of the computational process}, its \textit{output} and the \textit{code used} \cite{deMol}. 

Hence it seems that there is a need to revisit our current level of knowledge of the busy beaver problem with a view to providing sufficient evidence for the results known. In particular, there is a need for an appropriate methodology in which the results are not only determined, but also justified in sufficient detail to be checked and/or reproduced by an independent researcher.

There is also the issue of how to deal with the very large numbers involved. 
Indeed, without some \textit{prima facie} evidence that it is feasible to execute machines which produce such very large numbers, it is unclear that it is sensible to approach the problem at all. Even with modern computing power, we cannot hope to naively execute computations of the order of $10^{50}$ steps with any surety, let alone those in the order of $10^{30,000}$. Accordingly, we need to have some evidence that appropriate acceleration techniques are available so that such computations can be feasibly performed before spending time and energy pursuing them, especially as the busy beaver problem is to find the machines which produce the largest numbers. 

In this paper we address both of these issues. We discuss the appropriate framework in which to investigate the busy beaver problem, including the issue of the nature of the relevant evidence and a suitable way to provide it. 
We also show how a simple technique that we call the \textit{observant otter}
\footnote{The alliterative allure and literary legacy of the term \textit{busy beaver} has inspired a number of similar names. 
} 
heuristic can be added to the well-known method of \textit{macro machines} \cite{HM90}\footnote{An unintensionally alliterative name perhaps, but entirely consistent with the busy beaver naming convention. :-)}, and that this combination is sufficient to feasibly evaluate all known `large' machines (which we refer to as \textit{dreadful dragons}). 

It should be noted that the busy beaver problem is an endless one, in that it seems we can only ever know a finite number of its values, and that whatever the maximum value known is at any point in time, there will always be a slightly larger value for which the problem is unsolved. However, that should not deter us from determining as much of it as we may, despite some understandably pessimistic projections that we may never solve the problem for machines with as little as 6 states \cite{Dewdney}, let alone for 10 or 20. Given the recent history of the development of computing technology, it would seem wise to ``never say never'' about such specific projections.\footnote{As Treebeard, the oldest living creature in Tolkien's Middle-Earth puts it ``\textit{Never} is too long a word, even for me ''. -- J.R.R. Tolkien, `The Lord of the Rings', p.1016, Harper-Collins, 1991.} 

It should also be noted that the provision of evidence for the busy beaver problem may be used to investigate related problems. One common example is the variant of the busy beaver problem in which the non-blank characters must occur contiguously. This may mean that the final configuration must consists of a single string, none of whose characters are blank, 
 (or at least the `score' allocated to each machine is based on the largest number of contiguous non-blank characters in the final configuration, rather than the overall number of such characters). Another possibility is to consider the most complex `shapes', rather than simply the largest number of characters. For example, the machine referred to above which terminates after $10^{36,534}$ steps leaves a string on the tape of the form  

$$1(110110)^X 11$$

where $X$ is a number with 18,627 digits. Whilst the sheer size of this string is notable, it does not seem particularly complex, as its basic structure may be thought of as the shape below.

$$1(110)^* 11$$

This is arguably less complex than the shape below, which is produced by a machine which prints  `only' around $10^{881}$ digits

$$
1(01)^* 101010010100101
$$

and certainly less complex than the shape below (from a machine with 2 states and 5 symbols)

$$130111242422(42)^* $$

In other words, it seems interesting to contemplate variations on the busy beaver problem, and that these variations would be straightforward to solve once a complete list of terminating machines of a given size has been provided. 

It is also interesting to consider more comprehensive properties of the class of terminating machines, rather than simply various maxima. For example, it is well-known that there is a 5-state 2-symbol machine which halts after 47,176,870 steps leaving 4,098 1's on the tape (and so this machine has productivity 4,098, and what we will call \textit{activity} 47,176,870). What is perhaps less well-known is that there is another machine which produces the same number of 1's, but halts after `only' 11,798,826 steps (see Section 6 below), and once can argue that this is perhaps a more effective machine, in that it produces the same productivity in around a quarter of the activity. Furthermore, there a total of 6 5-state 2-symbol machines with productivities around this same value, with the two mentioned above of productivity 4098, two more with productivity 4097 and a further two with productivity 4096. However, the next highest known productivity is 1915, the next highest 1471 and then 501, which suggests that these six machines are rather unusual. Analysis of this phenomenon has concentrated on determining the productivity of a particular machine; however, it would seem that it is at least as interesting to determine why only this small number of machines has this behaviour. 

A related and similarly intriguing problem is the \textit{placid platypus} problem \cite{Harland06}\footnote{The platypus is an Australian monotreme, and is shy and retiring by nature.}, which may be thought of as a rough inverse to the busy beaver problem. The placid platypus problem is to determine the terminating Turing machine of the smallest size that can print a given number of non-blank characters on the blank input. Hence a Turing machine of size $n$ with productivity $k$ shows that the busy beaver value for $n$ is at least $k$, and that the placid platypus value for $k$ is at most $n$. More discussion on this and similar problems is in Section~\ref{sec:discussion}; for now, we note that the evidence for the busy beaver problem (and more particularly the data on the terminating Turing machines of up to a given size) can be used to solve some similar problems of interest. This is because the provision of such evidence will involve at least a documented list of all machines of up to a given size, together with the classification of each machine as terminating or non-terminating (on the blank input), and their final configurations. This will not only allow verification of empirical results by other researchers, but also for other properties of interest to be discovered and tested.

This paper is organised as follows. 
In Section~2 we discuss related work, and in Section~3 provide some terminology and definitions. 
In Section~4 we discuss the issues in addressing the busy beaver problem and describe our framework for doing so. 
In Section~5 we discuss execution methods for Turing machines, and in particular the observant otter heuristic. 
In Section~6 we present the results of evaluating this heuristic on 100 machines, which include all the dreadful dragons known. 
In Section~7 we discuss various issues that arise, and in Section~8 we present our conclusions and possibilities for further work. 

\section{Related Work}

The busy beaver function is defined as the maximum number 
of non-blank characters that is printed by a terminating $n$-state Turing machine on the blank input. This
function is often denoted as $\Sigma(n)$; in this paper we will use
the more intuitive notation of $bb(n)$. The number of non-blank characters printed by
a terminating machine is known as its {\em productivity} \cite{BBJ}. We introduce the term \textit{activity} to denote the number of steps taken by the machine to terminate on the blank input. A non-terminating machine may be considered as having activity of $\infty$. In the literature, the maximum activity for machines of size $n$ is often denoted as ${\cal S}(n)$; in this 
paper, we denote this function as $ff(n)$.\footnote{We call this function the {\em frantic frog.}} 

Rado introduced the busy beaver function in order to provide a specific example of a non-computable function rather than relying on the more abstract diagonalisation arguments \cite{Rado62}. He showed the function is non-computable by showing that it grows faster than any computable function. Shortly afterwards, Green \cite{Green64} gave some lower bounds for this function, as well as some example `monster machines' (including, notably, two of the very few examples of machines with 7 or more states, see machines 99 and 100 in Section~6). Lin and Rado were the first to produce some specific values for the busy beaver function itself \cite{LR65} for machines with 1, 2 or 3 states (and with 2 symbols). Brady later extended this work to include the 4-state case, which required a more sophisticated analysis of non-termination behaviour \cite{Brady83}. Similar results were obtained independently by Machlin and Stout \cite{MS90}. 

Marxen and Buntrock were the first to significantly investigate the 5-state and 6-state cases, as well as introducing the notion of macro machines \cite{HM90}. Their initial findings of 6-state dreadful dragons have been gradually superseded by various contributors over the years, usually in unpublished work. Such contributions have also involved dreadful dragons with more than 2 symbols, again usually in unpublished work, and often with spectacularly larger results than for the 2-symbol cases. For example, the most productive 3-state 2-symbol machine prints 6 1's; the best known 3-state 3-symbol machine prints 374,676,383 non-blanks. An excellent summation and history can be found at Marxen's web site \cite{Marxen}, and also in the work of Michel \cite{Michel93,Michel}. 

Lafitte and Papazian have provided what seems to be the most comprehensive examination of various classes of the busy beaver problem to date \cite{LP07}. Lafitte and Papazian discuss  detailed analyses of the 3-state 2-symbol and 2-state 3-symbol machines, and also less detailed but still significant analyses of the 4-state 2-symbol and 2-state 4-symbol machines as well as the only extant analysis (to the best of the author's knowledge) of the 3-state 3-symbol machines. They also point out that the 2-state 4-symbol case is the first one where seemingly `chaotic' execution patterns occur, particularly amongst the machines which do not terminate, which make this class a more difficult one to analyse than any of the previous cases. They also give some results which show that the maximal machines tend to have the property that the machine's activity is around the square of the machine's productivity, and observe that the $n$-state 2-symbol case is generally simpler to analyse than the 2-state $n$-symbol one. They also provide some classifications of non-terminating machines. 

There is generally less analysis of the class of machines in which in which the product of the number of states and symbols (which we refer to as the \textit{dimension} of the machine) is 9 or greater. Of these classes, the 5-state 2-symbol machines have been the most studied, and there seems to be an informal consensus that the maximal productivity in this class is 4,098. However, it seems fair to say that no sufficiently detailed search of this class of machines has been performed to feel secure in the claim that this is indeed the maximum. Kellett \cite{Kellett}, building on the earlier work of Ross \cite{Ross}, has performed a search like this for a variant of Turing machines that can only move or change the tape. The unclassified cases are reduced to 98 holdouts, and whilst the list of holdout machines is explicitly listed, the evidence for their non-termination is again by human inspection, particularly for the final 13 which are classified as non-terminating ``by our own visual deductive reasoning system'' (\cite{Kellett}, p.80). This is probably the most comprehensive analysis to date of this class of machines. Kellett also reports some results for 6-state machines, but without any analysis of the non-terminating machines. 

It is also curious to note that despite being the `quadruple' variant of Turing machines (i.e.\ each transition is a quadruple rather than a quintuple, reflecting the fact that each transition can either change the symbol on the tape or move the tape head, but not both), the maximal productivity is also 4,098. Whilst it is well-known that the quadruple and quintuple variants of Turing machines are equivalent from a computability perspective (i.e.\ for any machine in one class there is an equivalent machine in the other), it is not obvious that this equivalence is maintained when the number of states is restricted; in particular, it is not obvious that for any 5-state quintuple machine that there is an equivalent 5-state quadruple machine (the converse is straightforward). This means that it is appropriate, if perhaps a little conservative, to consider the quadruple and quintuple cases as separate problems. There are some quadruple machines of notable size on Marxen's web page, but it is also worth noting that the largest dreadful dragon machines known are all quintuple machines. 

There have been some efforts to provide automated methods for proving non-termination results for 5-state 2-symbol machines, including those of Georgiev \cite{Skelet} and Hertel \cite{Hertel09}.
Georgiev's prover is able to prove non-termination for all but 164 machines. Hertel's prover is similar, in that it has been able to classify all but 100 machines. This seems to indicate that for many non-termination techniques, there is a small `hard core' group of machines which seem highly resistant to them. 
Unfortunately, both of these efforts appear unfortunately to be incomplete, and no longer the subject of active research. 

The work of Wolfram \cite{Wolfram84,Wolfram} has a different motivation than ours, in that Wolfram is interested in finding sophisticated computational behaviour in a wide variety of automata, and not just Turing machines. It seems hard to overstate the massive amount of work that has gone into enumerating and classifying the vast numbers and types of automata during this process. Whilst his focus is very much broader than ours, the basic method is similar, in that it is a matter of systematically searching through large numbers of machines and classifying their behaviour. However, Wolfram is more concerned with identifying particularly complex computational behaviour, rather than completely classifying a particular class of machines. 

There has been some investigation into busy beaver functions on different types of Turing machines. As mentioned above, the quadruple variant is one. Batfai \cite{Batfai09b} has also shown how introducing a `skip' action (i.e.\ introducing the option that a machine does not have to move the tape head during a transition) can generate machines of greater activity (as well as not having an explicit halting state). There has also been an investigation of busy beaver functions on a one-way tape \cite{Walsh} rather than a two-way one.

Many presentations of Turing machines assume that there is a single semi-infinite tape. In this paper, as is standard with busy beaver investigations, we assume that there is a single tape which is infinite in both directions. Whilst the difference between a semi-infinite tape and one that is infinite in both directions is moot from the perspective of computability \cite{Sudkamp}, the fact that it is impossible to hit the end of the tape simplifies some matters with the analysis of the execution of machines, and in particular that there is no possibility of `abnormal' termination by coming to the end of the tape. 

The extension to allow two-dimensional tapes provides a significant increase in complexity and possibilities. Brady called these ``TurNing Machines'' \cite{Brady88} and investigated many of their properties. Independently, Langton proposed what he called \textit{ants} \cite{Langton86}, which turned out to be conceptually similar, and  were shown later to be computationally universal \cite{GMG02}. As is not uncommon, similar ideas occurred to others as well, and were called ``Tur mites'' \cite{Dewdney89}. There are some fascinating developments here, including interesting variations on the geometry of the ``tape'' allowed \cite{RTR}. Similar ideas can also be found in Wolfram's work \cite{Wolfram}. Whilst this is an intriguing area for investigation, in this paper we focus on machines with a single one-dimensional tape. Whilst the difference between a single semi-infinite tape, one that is infinite in both directions, and a two-dimensional is moot from the perspective of computability \cite{Sudkamp}, the use of a single infinite tape not only eliminates some possibilities for `abnormal' termination (by coming to the end of the tape), it also provides a simpler context and ease of comparison with existing results. 

The busy beaver function is often denoted as $\Sigma(n)$ where $n$ is the number of states. Strictly speaking, this should be $\Sigma(n,m,d)$ where $n$ is the number of states, $m$ is the number of symbols and $d$ is this number of directions in which the tape head can move (which reflects the dimensionality of the tape). For a one-dimensional tape, $d$ is 2, and as this will always be the case for the machines considered in this paper, we will only consider the number of state and number of symbols. We will also use the more intuitive notation of $bb(n,m)$ for productivity, and similarly $ff(n,m)$ for activity. 

The main use of this function has been as a simple example of
non-computability, especially as there are also larger functions which
can be defined similarly.  A notable use of this function is given in Boolos, Burgess 
and Jeffrey \cite{BBJ}, in which the busy beaver function is the
subject of the first undecidability result established, rather than
the more usual choice of the halting problem for Turing machines.

The relationship between $bb(n,2)$ and $ff(n,2)$ has been investigated
\cite{Julstrom,BJZ96,BP02}, and it is known that $ff(n,2) < bb(3n+c,2)$ for a constant
$c$ \cite{YDX}. However, this is still rather loose, and does not give
us much insight into the relationship between $bb(n,2)$ and $ff(n,2)$. 
In a similar manner, lower bounds on $bb(n,2)$ have been known for some
time \cite{Green64}; however, those given for $n \leq 6$ have been far
surpassed already. 

As mentioned above, Marxen and Buntrock \cite{HM90} were the first to introduce the notion of macro machines, which is significant for being the first concrete improvement over the naive implementation of a Turing machine. This approach significantly increases the efficiency of execution of Turing machines, and has been used as the basis for further improvements by Holkner \cite{Holkner04} and Wood \cite{Wood08}. The approach used in this paper is largely based on these two papers. 

\section{Terminology and Definitions}

Before discussing our framework and results, we make some of our terminology and definitions precise. 

We use the following definition of a Turing machine \cite{Sudkamp}.

\begin{definition}
A Turing machine is a quadruple $(Q \cup \{z\}, \Gamma, \delta, a)$ where

\begin{itemize}
\item $z$ is a distinguished state called a {\em halting state}
\item $\Gamma$ is the tape alphabet
\item $\delta$ is a partial function from $Q \times \Gamma$ to $Q \cup \{z\} \times \Gamma \times \{l,r\}$ called the {\em transition function}
\item $a \in Q$ is a distinguished state called the {\em start state}
\end{itemize}
\end{definition}

Note that due to the way that $\delta$ is defined, this is the so-called \textit{quintuple transition variation} of
Turing machines, in that a transition must specify for a given input state and input character, a new state, an output character and a
direction for the tape in which to move. Hence a transition can be specified by a quintuple of the form

\begin{center}
$(State, Input, Output, Direction, NewState)$
\end{center}
where $State \in Q$, $NewState \in Q \cup \{z\}$, $Input, Output \in \Gamma$ and $Direction \in \{l,r\}$. 

We call a transition a \textit{halting transition} if $NewState = z$; otherwise it is a \textit{standard transition}. 

Given some notational convention for identifying the start state and halting state, a Turing machine can be characterised by the
tuples which make up the definition of $\delta$.

Note that there are no transitions for state $z$, and that as $\delta$ is a partial function, there is at most one transition for a given
pair of a state and a character in the tape alphabet. This means that our Turing machines are all \textit{deterministic}; there is never a case when a machine has to choose between two possible transitions for a given state and input symbol. Note that it is possible for there to be no transition for a given state and symbol combination. If such a combination is encountered during execution, the machine halts.  We generally prefer to have explicit halting transitions rather than have the machine halt in this way. 

We denote by an $n$-state Turing machine one in which $|Q| = n$. In other words, an $n$-state Turing machine has $n$ standard states and a halting state.  

We will find the following notion of \textit{dimension} useful.  

\begin{definition}
Let $M$ be a Turing machine with $n$ states and $m$ symbols (where $n,m \geq 2$). Then we say $M$ has \textit{dimension} $n \times m$. 
\end{definition}

A configuration of a Turing machine is the current
state of execution, containing the current tape contents, the current
machine state and the position of the tape head \cite{Sudkamp}. We will use
$111\{b\}011$ to denote a configuration in which the Turing machine is
in state $b$ with the string 111011 on the tape and the tape head
pointing at the 0.

Machine states are labelled $a,b,c,d,e \ldots$ where $a$ is the initial state of the machine. The halting state is labelled $z$. Symbols are labelled $0,1,2,3,\ldots $ where $0$ is the blank symbol. 

We will use the term \textit{hops} to refer to the number of steps needed to execute the machine in a naive manner. Clearly many of the machines in Table~\ref{table1} can never be feasibly executed in this way. More sophisticated methods (see Section~\ref{sec:execution} below) are able to determine particular configurations without executing every intermediate one. We will use the term \textit{steps} to refer to the number of execution steps in an implementation; each step in an implementation may correspond to a very large number of hops. Clearly a larger the ratio of hops to steps will provide greater acceleration in execution time than a smaller one, and as we shall see, this ratio may vary significantly between machines. 

\section{Framework}

The process for solving an instance of the busy beaver problem or a similar problem involves the four steps below.

\begin{enumerate}
	\item Select a precise definition of the machines to be evaluated. 
	\item Enumerate all appropriate machines.
	\item Classify the machines into terminating and non-terminating.
	\item Search the terminating machines to determine the appropriate property. 
\end{enumerate}

The output from Step 1 is a particular definition of a class of machines. 
The output from Step 2 is an appropriately formatted file (or number of files) containing a numbered list of all machines enumerated, as well as evidence for the exclusion of any machines not enumerated. We discuss this in more detail below.
The output from Step 3 is the list from Step 2 augmented with the classification of each machine as terminating or non-terminating. For terminating machines, this should also include the final configuration and the corresponding number of hops. For non-terminating machines, this should also include some evidence for the non-termination, such as an infinitely increasing pattern. 
The output from Step 4 will depend on the nature of the property involved. For the busy beaver problem, this should include the busy beaver value (i.e.\ the maximum productivity), the definition of the relevant machine, the final configuration and the machine's activity. In many cases, the machine with the maximum productivity is also the machine with the maximum activity; if this is not the case, the same information should also be given for the machine with the maximum activity (we will call this the \textit{frantic frog} machine). The only known case so far in which the busy beaver machine differs from the frantic frog machine is for machines with 3 states and 2 symbols. 

The output from Steps 2, 3 and 4 should also include all code used. 

This may sound simple enough in principle, but as in many problems, the devil is in the detail.\footnote{Can we call instances of this phenomenon \textit{detail devils}? :-)}

Step 1 may seem trivial but is fundamental to this problem. It is known that there are many varieties of Turing machines, which, from the perspective of computability, are all equivalent. However, in this context the size of the machines are deliberately limited, which means that variations may have a significant effect. In much of previous literature on the busy beaver problem \cite{Rado62,LR65,Brady83} the machines have one tape which is infinite in both directions. However, there is no reason that a one-way tape cannot be used \cite{Walsh}. As discussed above, there are other variations as well which include quadruple machines (i.e.\ those which can only move or change the symbol on the tape in a transition, and not both), machines which may choose to not move on some steps and multiple tapes or two-dimensional tapes. Even once such a definition is decided upon, there are still some decisions to be made. For example, the existing known dreadful dragons all have a single halt transition which is explicitly defined in the machine. Not all varieties of Turing machine have this property; some signal halting when there is no defined transition for the current state and input symbol rather than having an explicit halting state, and others do not terminate on any inputs (Wolfram's Turing machines are one such variant \cite{Wolfram}). 

All the known dreadful dragon machines except one (see Section~6 below) are quintuple machines with exactly one halting transition, and all other transitions defined. In other words, these are machines with $n$ states and $m$ symbols, one halting transition and exactly $(n \times m) - 1$ standard transitions. In principle, there is no obvious reason why a dreadful dragon must have exactly $(n \times m) - 1$ standard transitions (although this seems intuitively plausible), and so we should allow for the possibility that it may have less than this number. This corresponds to having more than one halt transition, or a number of undefined transitions. For candidate busy beaver machines, it seems natural to insist on explicit definition of transitions, so that a machine will only halt when a halting transition is executed, rather than halting due to having a missing transition. This may mean that there are multiple halting transitions, but it also means that the extra execution step for the halting transition may lead to an increase in productivity, as if the output symbol from a halting transition is always non-blank, then executing such a step can never reduce productivity, and may increase it. Hence for our instance of the busy beaver problem, we insist that there be at least one halting transition, but not that there be exactly one such transition. There may of course be reasons to vary this in some other contexts; the main point to note is that it is necessary (if seemingly pedantic) to explicitly state assumptions such as these.  

Step 2 seems simple enough; once the decision in Step 1 is made, the definition of whichever variety of Turing machines should make it straightforward to generate all such machines, and record the outcome. In practice, there is a very large number of such machines, making optimisation a critical issue. Specifically, for a machine of dimension $k$, there are $O(k^k)$ possible machines. This `hyperfactorial'\footnote{Just as $2^n + n!$ is O$(n!)$, $n! + n^n$ is O$(n^n)$, hence the name.} growth rate is one that cannot be taken lightly, except perhaps in some of the smaller cases.

In general, we are interested in evaluating the busy beaver and similar properties for machine of a specific dimension. This makes it seem natural to start with the smallest dimensions, and incrementally generate machines of the larger dimensions.  
Chaitin has noted that separating machines into classes by the number of states (and hence by dimension) is not the best way stratify the overall class of machines, and that it is better to do so by the number bits required to represent each machine \cite{Chaitin87}. Whilst this seems perfectly reasonable, we have chosen to stick with the process of classifying machine classes by the number of states and symbols in order to allow direct comparisons with previous work. However, once the generation is done, it would relatively simple to reorganise the machines to reflect such an ordering.  

An important observation on the generation process is that we will assume that as the machine commences execution in state $a$, the second state encountered we will label $b$, the third  $c$, and so on (and similar for symbols with the first non-blank being $1$, the second $2$ and so on). This is to ensure that we do not generate machines which are identical apart from the names of the states involved. For example, if we consider the set of possible states as say $a,b,c$ and symbols $0,1,2$, and then generate all possible machines with these names, then we will generate a number of machines which only differ by a trivial renaming of states or symbols. For example, the two (partial) machines below behave identically, apart from the names of the states $b$ and $c$, and one can be obtained from the other just by swapping these two names.

\begin{center}
$\{(a,0,1,r,b), (b,0,1,l,c), ...\}$ \\
$\{(a,0,1,r,c), (c,0,1,l,b), ...\}$ 
\end{center}

Similarly, if the first two transitions were $(a,0,2,r,b)$ and $(b,0,3,l,a)$ then by renaming $2$ as $1$ and $3$ as $2$ we will get an identical machine except for this swapping of symbol names. 

It is also important to keep in mind that for the busy beaver and variants of it, these machines will only ever be executed with the blank input. This makes it simple to identify some classes of machine that will be uninteresting. For example, any machine in which the first transition is also a halting transition is of no interest. Additionally, there is a large number of machines for which it is trivial to show non-termination. For example, as observed by Lin and Rado \cite{LR65}, if we denote the starting state of the machine as $a$, then if the machine contains a transition of the form $(a, 0, \_, \_, a)$, (i.e.\ for the input state $a$ and input symbol $0$, the output state is $a$), then the machine will never halt on the blank input, no matter what choice of output symbol or direction is made. Hence we need only consider machines for which the output state for the transition for state $a$ and symbol $0$ is not $a$. Furthermore, as discussed above, we can assume that this state is $b$. 

This means that the first transition will always be of the form $(a,0,\_,\_,b)$\footnote{We use the notation $\_$ to indicate an arbitrary choice.}. The choice of the direction for the first transition is arbitrary, as the tape is infinite in both directions. Historically, there have been choices made for either direction; in our case, we have chosen for the machine to move to the right.\footnote{This is in line with a number of previous authors, although Green\cite{Green64} made the opposite choice. Note that this means that for any machine, there is a \textit{sinister sibling} in which all the directions in all transitions are reversed, and which will have the same productivity and activity as the \textit{orthodox original}.} Finally, note that there is no point in having the initial transition with the output symbol being blank. As the tape is initially blank, this only serves to change the state of the machine. If the machine never prints any non-blank character, then it is clearly of no interest (whether it terminates or not). Otherwise, we can identify the first state in which a non-blank character is printed, say $c$, in which case we can then swap the states $a$ and $c$ to produce a machine whose first transition has the output symbol being a non-blank. As discussed above, we can assume this is $1$ (the first non-blank symbol). This means that the first transition in any machine of interest will always be $(a,0,1,r,b)$. 

We can use similar reasoning to eliminate machines containing a transition of the form $(b,0,\_,r,a)$ or of the form $(b,0,\_,r,b)$, as in both cases the machine will move infinitely to the right when commencing on the blank input. Hence the second transition must be one of the following forms: $(b,0,\_,l,a)$, $(b,0,\_,l,b)$ or $(b,0,\_,\_,c)$.\footnote{This last case is clearly only applicable when there are at least three states allowed.} Note that this specifically excludes this second transition being a halting transition, as clearly machines which only execute two steps are of no interest. 

It should be noted that there are two considerations at work here. One is to eliminate duplication by generating only one machine from each equivalence class whenever possible. This is the principle behind the restriction that the second state is always $b$ and the first non-blank symbol $1$, as this ensures we do not generate multiple machines which can be shown to be equivalent by renaming states or non-blank symbols. The other is to only eliminate machines which are of no interest, either because they terminate trivially (such as on the second step of computation) or will obviously not terminate (such as moving infinitely to the right). Clearly if there is any doubt about a particular machine, we must include it in the list of those generated, as otherwise it will not be considered any further. In other words, we must have evidence that any machine excluded from the list generated is known to be irrelevant for the property under consideration. For the busy beaver problem, it is clear that any machine that terminates in one or two steps is irrelevant, as is any machine which moves simply moves infinitely in one direction in the above manner. In fact, it is difficult to think of any desired property for which machines with such trivial computational properties will be of interest.\footnote{Any machine which does not terminate is irrelevant to the busy beaver problem. However, it may be that more complex types of non-terminating machine may have properties worthy of study. It is clear, though, that the non-terminating machines excluded by the above process are not of interest.}

A final consideration is that because we are only interested in executing these machines with the blank input, we can further reduce the number of machines generated, by means of a process that has been called \textit{tree normal form} \cite{LR65}.
The idea is that rather than generate all possible machines, we need only consider machines generated by executing a partially defined machine on the blank input (starting with just the initial transition $(a,0,1,r,b)$) until we come across a combination of input state and input symbol for which there is no transition defined in the machine. We then generate a transition (possibly selecting one from a number of alternatives), add it to the machine and keep executing. This process will continue until either the (partial) machine is shown not to terminate, or we choose to add the halting transition, at which point the machine definition is complete. We can then output the machine (such as writing it to a file), and then backtrack over the most recent choice to find a different alternative, and proceed. Hence this process is a large backtracking search over the possible ways in which a machine of the given dimension could execute. 

The circumstances under which we choose to add the halting transition are when we know that sufficient states and symbols are already present to meet the requirements of the machine dimensions. For example, if we are generating all machines with 3 states and 3 symbols, then we must have states $a$, $b$ and $c$ already present in some transition, as well as the symbols $0$, $1$ and $2$. Otherwise, we will have generated a machine of a smaller dimension than the required one. It should also be noted that once we reach this point, the current computation does not necessarily terminate, either at this step or at all. As above, if we are generating machines of dimension 9 and we have already defined 7 transitions, then when we determine the 8th one, the one remaining unspecified transition must be a halting transition. Given that the machine is now completely specified, we output the machine, and then backtrack over the most recent choice, but at this point the computation has not reached the halting transition (and possibly may never do so).

Note also that we may choose to add the halting transition earlier than the last remaining transition. For example, if we have the partial machine below when generating a machine of dimension 9

$$(a,0,1,r,b),(b,0,1,r,c), (c,0,2,l,a)$$

\noindent
then as all the states and symbols are present, we may choose to add the halting transition $(a,1,1,r,z)$, even though this is less than the total possible number of transitions for machines of this dimension. However, as discussed earlier, we cannot be certain that this machine is of no interest, and so we include it. 

The tree normal form process means that the number of machines needed to be stored is considerably reduced; for example, Machlin and Stout's implementation for the 4-state 2-symbol machines produced `only' 603,712 machines rather then 25,600,000,000 \cite{MS90}. This is certainly welcome, due to the hyperfactorial number of machines of a given dimension (as noted above). However, this does complicate the process, as it tends to blur Steps 2 and 3, and in particular requires at least a basic machine execution engine, ideally together with some non-termination tests, in order to generate the machines to be classified in the first place. This process of minimising the number of machines generated by merging machine generation and execution is presumably why it has been hard to find reproducable evidence from previous searches (together with the amount of information that needs to be stored, which in previous years was significantly more problematic than now). However, it is our belief that by appropriately limiting the amount of time spend executing during this process and outputting machines as they are generated, it is possible to both minimise the number of machines generated and store the results of the generation process for subsequent processing. 
 
One observation of note is that for machines of a given dimension, the number of (all) machines in each class is same. For example, considered naively, there are the same number of 5-state 2-symbol machines as 2-state 5-symbol ones. To see this, consider that each machine with $n$ states and $m$ symbols will have $n \times m$ transitions of the form $(State, Input, Output, Direction, NewState)$, one for each combination of $State$ and $Input$. For each such transition, there are $n \times m \times 2$ possible combinations of $Output$, $Direction$ and $NewState$.\footnote{This is a slightly inaccurate estimate, as we need to allow for (at least) one such combination to be a halting state. But whatever approach to halting transitions is used, each class will contain the same number of machines.} When generating machines via the tree normal form process, this equivalence does not hold in principle, but in practice the number of machines in each class tends to be comparable. In general there will be slightly less 2-state $n$-symbol machines than $n$-state 2-symbol ones due to their only being two possibilities for the $b,0$ transition rather then three. 

It should also be noted that this separation of the classification phase into a preliminary execution phase (i.e.\ to define the machines) and then a more elaborate execution phase (if needed) will significantly enhance both the development of increasingly powerful methods for classifying machines and possibilities for collaboration between interested researchers. The reason for this is that it seems inevitable that each new class of machines generated will require a number of iterative refinements to execution methods in order to classify them as terminating or non-terminating (echoing the discussion of de Mol\cite{deMol}). In other words, it seems far-fetched at best to expect that the first time that machines of say dimension 16 are generated that the methods developed for smaller methods will classify all such machines. Presumably there will some simple cases (and quite possibly a large majority) which are able to be classified, but there will undoubtedly be a significant fraction which will require methods of greater sophistication than seen previously. Hence, once the machines are first generated, we would expect to identify an initial set of unclassified machines, which would then be gradually reduced (hopefully to zero) by a period of observation, development, experimentation and refinement. It also seems prudent to be able to distribute a complete generation but partial classification, to allow for collaboration with other interested researchers, or to conduct parallel analyses on different platforms. This consideration will be increasingly important as the dimensions (and hence number of machines) gets larger.  In this way it seems only practical to have this separation between the generation process and the classification one, despite the commonality between them due to their common basis in machine execution. 

Step 3 is the one that poses the most technical challenges. It is well-known that the problem of determining whether an (arbitrary) Turing machine halts on the blank input is undecidable \cite{Sudkamp}. This would make it seem that attempting to automate the process of classifying such machines is doomed to failure, and that there will always be a `hard core' of machines which will resist all attempts at automated classification. However, this is not strictly the case. Whilst we cannot hope to have a single classification algorithm for all Turing machines, it is important to note that we are only ever interested in a specific finite set of Turing machines (i.e\ those of the current dimension of interest), and, strictly speaking, the problem of classifying a finite set of Turing machines into terminating and non-terminating on the blank input is decidable. The key point here is that as there are only finite number of machines under consideration (say $n$), then there are only $2^n$ possible classifications.
For any one of these classifications, we can construct a Turing machine which will take as input one of the $n$ machines and output the correct answer according to the classification. Naturally we have no way of knowing \textit{a priori} which of these $2^n$ machines is the correct classifier, but this machine certainly exists.  Whilst the existence of this machine means that the problem is decidable, this knowledge is of little or no use when it comes to actually finding the appropriate classification algorithm. However it does mean that it is not inevitable that there must be a `hard core' as mentioned above for all such machine dimensions. 

It is also worth noting that the well-known proof of the undecidability of the halting problem for Turing machines on the blank input involves machines of an unbounded size. The most common reduction of the halting problem for Turing machines with input $w$ to the halting problem for Turing machines on the blank input involves constructing a machine $M'$ which will take the blank input, print $w$ on the tape and then terminate, and then prepending this machine $M'$ to the original machine $M$, giving a new machine $M''$ \cite{Sudkamp}. If we then run $M''$ commencing with the blank tape, this will first print $w$ on the tape, and then proceed exactly as $M$ would on input $w$ (provided that we have appropriately arranged the connection between $M'$ and $M$). \textbf{The key point to note is that both $w$ and $M''$ are of unbounded size.} As for the busy beaver problem we are only enumerating machines (equivalent to $M''$) of up to a given size, we can only ever `simulate' (using the above transformation as a template) not only a Turing machine $M$ of a bounded size but also an input $w$ of bounded size. Hence by attempting to find a classification algorithm for machines of a specific dimension we are not attempting the impossible task of finding an algorithm for an undecidable problem, but for a finite (and hence decidable) `underapproximation' of this problem. In other words, the undecidability of the halting problem for Turing machines on the blank input means that there is no single algorithm which will correctly classify all Turing machines; however, this leaves open the possibility that for all Turing machines of a specific dimension, there is an algorithm which will correctly classify all machines in this class. This means that the process of finding classification algorithms will be a kind of `G\"{o}delian arms race', in that for any machine dimension with a classification algorithm $A$, there will always be larger machine dimension for which $A$ will not work, but that this larger dimension will have some classification algorithm, say $B$, which in turn will fail on some larger dimension, and so on. 

In summary, Step 3 is decidable, and hence not formally hopeless. However, it is still very much a significant practical problem to determine the appropriate classification algorithm. 

Step 4 is clearly the simplest of these steps. For the busy beaver, once we have a classification of all the machines output in Step 2, it is then a simple matter to find the maximum number of non-blank characters in the final configuration of all the terminating machines. There may also be many other variations that could be determined at this point, such as the maximum number of non-blank characters for terminating machines in which all non-blank characters are contiguous in the final configuration, (or possibly requiring the same constraint on all configurations in the trace) or the maximum number of contiguous non-blank characters in the final configuration, whether or not all non-blanks occur contiguously or not. It may also be of interest to examine the non-terminating machines, such as finding further examples of Brady's ``tail-eating dragons'' \cite{Brady83}. Naturally we need to ensure that any information required in this step is provided by the output from Step 3. 

Given that we need an execution method that can work with partially defined machines in order to perform Step 2, it seems natural to commence an investigation with the development of execution methods (for executing partially defined machines, for dealing with dreadful dragons, and possibly also with at least some simple non-termination checks) before considering the generation process in more detail. It seems reasonable to expect that the execution method will need to be continually improved and refined as the dimension of the machines explored rises, whereas we would generally expect the generation method to remain the same for any dimension. 

As noted above, for each class of machine of the same dimension there are approximately the same number of machines. For this reason it seems appropriate to approach this problem one dimension at a time, rather then one state at a time. We use the categorisation below. 

\begin{description}
\item[Blue Bilby:]\footnote{A bilby is a small cute Australian marsupial.} This includes machines of dimension up to 6, and hence covers 2-state 2-symbol, 3-state 2-symbol and 2-state 3-symbol machines. These are all cases for which the busy beaver results are known, and for which the number of machines is not enormous. In fact, it seems reasonable to be able to enumerate all machines in this category, whether arbitrary machines, those which conform to the above restrictions on the first and second transitions, and those generated by the tree normal form process. We refer to these three types of enumeration as \textit{all}, \textit{free} and \textit{tnf} respectively.  These can be used as test cases for larger categories, as well as for exploring the effectiveness of the tnf process. 

The bilby is a small and cuddly animal, reflecting that this category does not contain any monster machines.

\item[Ebony Elephant:] This covers machines of dimension 8, i.e.\ 4-state 2-symbol and 2-state 4-symbol machines. As noted above, machines in this category start to pass the limits of computations that can easily be performed or checked by hand, particularly in the 2-state 4-symbol case. It is likely the generation of all machines will be impractical, but free generation may still be feasible. 

The elephant is of course a creature larger than humans, but only poses danger under certain circumstances. 

\item[White Whale:] This covers machines of dimension 9 and 10, i.e.\ 3-state 2-symbol, 5-state 2-symbol and 2-state 5-symbol machines. This category is generally significantly beyond human computation, and for which only tnf generation seems feasible. This is also likely to be the largest category that can be feasibly explored on commodity hardware. 

The whale is a creature which should not be approached by humans without caution and a significant level of technological sophistication. 

\item[Demon Duck of Doom:]\footnote{This is a colourful nickname for an enormous flightless bird that lived in prehistoric Australia.} This category covers machines of dimension 12, which means 6-state 2-symbol, 4-state 3-symbol, 3-state 4-symbol and 2-state 6-symbol machines. This is not only the largest category considered so far, both in the number of machines and the number of different classes of machine, but also one that will be significantly more difficult to explore, and the one for which we have only very limited information at present. Given the numbers of machines, it is likely that this will only be feasibly explored via cloud computing methods (i.e.\ a distributed network of co-operating processors), rather than on a single desktop or similar machine. 

The Demon Duck of Doom, or Thunderbird, stood over 2.5 metres tall and lived some 15 million years ago, reflecting that the machines in this category are largely unknown, are likely to contain complexities that we have never seen before and cannot really be approached at all at present.
\end{description}

It seems premature at this point to name any larger categories, due to the significant amount of effort required to settle questions of interest for machines of dimension 12, let alone further dimensions. 

It is our intention to solve the busy beaver problem and various related problems for all of the above categories, including the Demon Duck of Doom. This will require a large amount of work (and an unbounded amount of time) on methods for the generation, analysis and execution of the various classes of machine above. It is likely that the classes up to and including the White Whale can be resolved using a single desktop machine; however, the sheer scale of the problem for the Demon Duck of Doom will almost certainly require co-operative methods for all aspects of the problem. 

It would also seem that the development of execution and analysis methods will be an ongoing process of development and refinement, even for the comparatively small cases such as the Ebony Elephant. In what follows we begin this process by seeing what can be done for executing machines such as the dreadful dragons given in Table~\ref{table1}.
 
\section{Executing Machines}\label{sec:execution}

\subsection{Macro Machines}

Turing machines are conceptually simple, particularly when it comes to executing them. 
Macro machines \cite{HM90} is a well-known technique for optimising the execution of Turing machines. Essentially, the tape is considered to consist of blocks of characters of a given fixed length $k$, and the transitions for the machine are determined according to this structure. For example, if $k=3$, then when the machine encounters say the input string $001$ in state $c$, then it will execute naively until either the tape head moves outside these three specific symbols, or it repeats an earlier configuration. In the latter case, we may classify the computation as non-terminating. Otherwise, we note the state in which the machine exits from the area containing just the three given symbols, and the symbols that now occupy these places, and consider this a new `macro' transition for the machine. The only additional information that is necessary is to keep track of the direction from which the tape head entered the current string, and the direction that it exited from the string. This means that for a given string, there are two transitions: one for when it enters from the left, and another for when it enters from the right. This also means that we need to keep track of not just the current (macro) symbol, but the current orientation of the tape head (left or right). Extending the notation for configurations above, we will use $111 \{b\} 001 \{l\} 110$ to denote a configuration in which the machine is in state $b$ with the string 111001110 on the tape, and the tape head pointing to the left hand end of the 001 string.  

For our earlier example in which the machine encounters the string $001$ in state $c$, if the initial symbols $001$ are converted to $111$ and the tape head leaves these three symbols in the state $d$, then we may consider this a macro transition from state $c$ with input $001$ to new state $d$ with output $111$. The details are described in Marxen and Buntrock's paper \cite{HM90}, and there are various improvements described on Marxen's web site \cite{Marxen}. 

A key aspect of macro machines is that having a fixed string length makes it easy to represent repetitive sequences of strings such as $001001001001$ as $(001)^4$. This, together with the acceleration technique below, is what makes macro machines significantly more efficient than naive execution.

To see how this works, consider a macro machine with $k=3$, and an execution trace in which the current state is $b$, the current string under the tape head is $(001)^{12}$, and the input direction is left (so that the tape head is pointing at the first $0$ in the string $(001)^{12}$). If the output state is also $b$, the output string is $111$ and the output direction is right (so that the tape head exits the string at the right-hand end), then it is clear that a machine configuration of the form 

$$X \{b\} (001)^{12} \{l\}Y_1 Y_2$$

\noindent will become 

$$X (111)^{12}  \{b\} Y_1 \{l\} Y_2$$

\noindent where $Y_1$ has length 3. 

If the output state is not $b$ (i.e.\ it is different to the input state) or the output direction is not to the right, then we ``decouple'' one copy of $001$ from the longer string $(001)^{12}$ and proceed. In the above example, if the output string is $110$, the output state is $c$ and the output direction is left, then the configuration 

$$X_1 X_2 \{b\} (001)^{12} \{l\} Y$$

\noindent will become 

$$  X_1 \{c\} X_2 \{r\} 110 (001)^{11} Y$$

\noindent where $X_2$ has length 3. 

Hence the two basic steps in the execution of a macro machine are those below.

\begin{itemize}
	\item If the input and output states are the same, and the input and output directions are opposite, apply the acceleration technique described above.
	\item Otherwise, update the configuration for one macro symbol only. 
\end{itemize}

Once the length of the strings is determined, it is possible to generate the definition of the macro machine by considering all $m^k$ such strings (for a $m$-symbol machine) and finding the corresponding macro transitions. A better approach is to generate them lazily during execution, i.e.\ if the current macro character under consideration does not have a macro transition defined for it, we calculate the transition as above and add it to the machine. This means that only the macro transitions that are actually needed in the machine are generated. 

Further discussion on this and similar points can be found in Marxen and Buntrock's paper \cite{HM90}. There are further optimisations that can be made for macro machines, but as our focus is on the observant otter heuristic and its effects, we only use this simple form of macro machines rather than the more sophisticated mechanisms. 

\subsection{The Observant Otter}

\begin{figure}
	\centering
		\includegraphics[width=15cm]{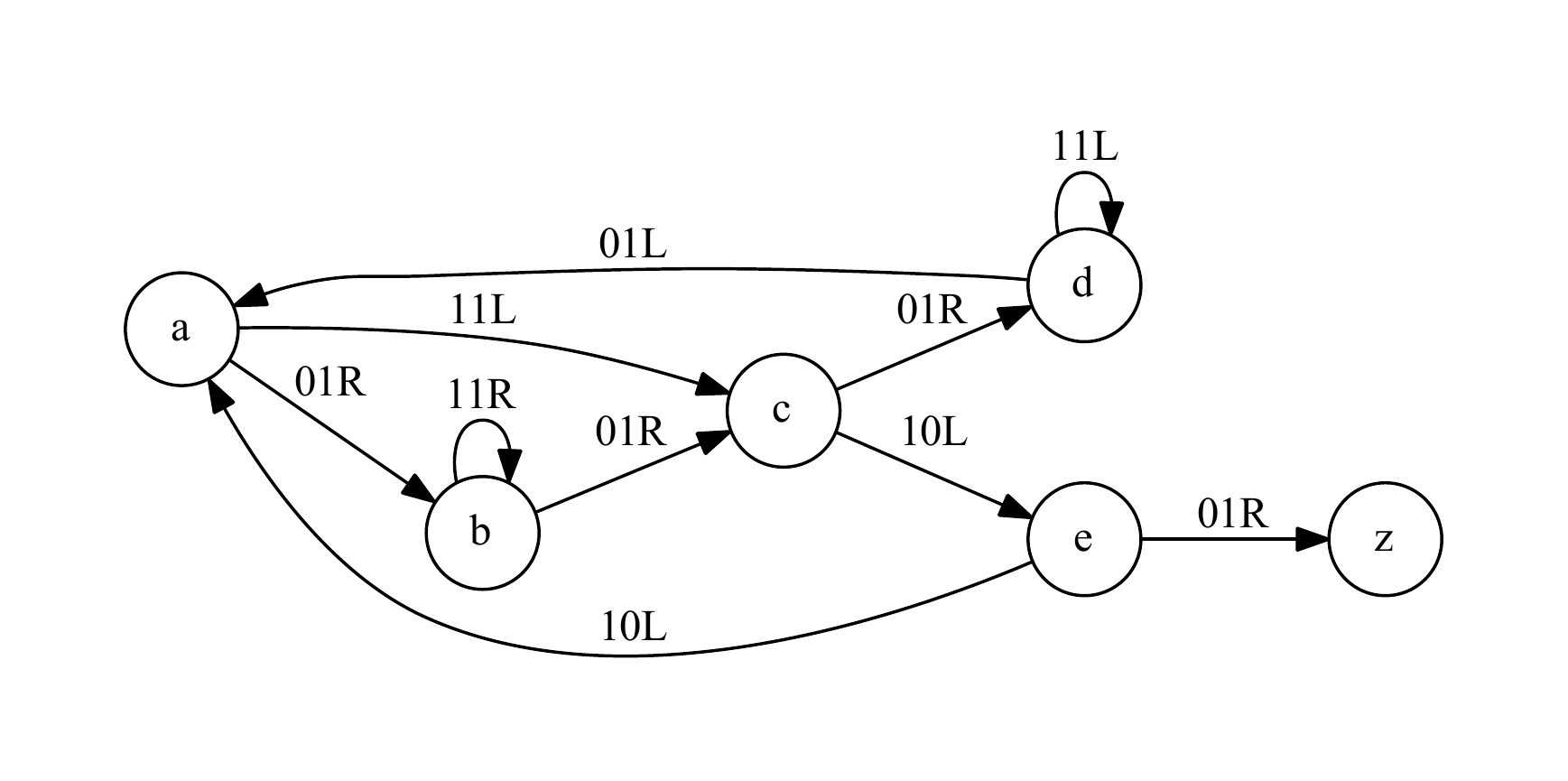}
	\caption{Best 5x2 candidate known}
		\label{fig:bb5}
\end{figure}

Consider the Turing machine in Figure~\ref{fig:bb5}. This is the current ``champion'' machine with 5 states and 2 symbols, which has activity 47,176,870 and productivity 4,098.\footnote{It is also arguably the most studied machine in the busy beaver literature.}

During the execution of this machine, the following configurations occur.

\begin{center}
\begin{tabular}{ll}
\textbf{Hop} & \textbf{Configuration }\\
12393	& $001(111)\{b\}(001)^{66}100$ \\
12480 & $001(111)^{6}\{b\}(001)^{63}100$ \\
12657	& $001(111)^{11}\{b\}(001)^{60}100$ \\
\end{tabular}
\end{center}

Clearly there is a pattern here, which is that a configuration of the form 

$$001(111)^{X}\{b\}(001)^{Y}100$$
\noindent becomes 
$$001(111)^{X+5}\{b\}(001)^{Y-3}100$$
\noindent provided that $Y > 3$. Specifically there will be 19 further occurrences of configurations of this form, as below.

\begin{center}
\begin{tabular}{ll}
\textbf{Hop} & \textbf{Configuration }\\
12924	& $001(111)^{16}\{b\}(001)^{57}100$ \\
13281 & $001(111)^{21}\{b\}(001)^{54}100$ \\
13728	& $001(111)^{26}\{b\}(001)^{51}100$ \\
...   & ... \\
29436	& $001(111)^{96}\{b\}(001)^{9}100$ \\
31233	& $001(111)^{101}\{b\}(001)^{6}100$ \\
33120	& $001(111)^{106}\{b\}(001)^{3}100$ \\
\end{tabular}
\end{center}

It is clear that whilst the macro machine method will produce all of these configurations, there will be considerable savings if we can predict the final occurrence of this pattern from the first few occurrences. In this case if we can use the first three instances of the pattern (those at hops 12393, 12480 and 12657) to predict the rest of the sequence, then we can jump from hop 12657 directly to hop 33120, thus advancing the hop count by over 20,000 in one single step of computation. 
 
Similar patterns, with increasingly large numbers, recur throughout the execution of this machine. For example, the following three configurations occur later in the computation. 

\begin{center}
\begin{tabular}{ll}
\textbf{Hop} & \textbf{Configuration }\\
281384 & $001(111)\{b\}(001)^{316}100$ \\
281471 & $001(111)^{6}\{b\}(001)^{313}100$ \\
281648 & $001(111)^{11}\{b\}(001)^{310}100$ 	 \\ 
\end{tabular}
\end{center}

As above, we can use this sequence to predict a configuration of 

$$001(111)^{526}\{b\}(001)^{1}100$$

The effectiveness of this approach seems to increase exponentially (at least for this example machine) as the computation goes on. Out of a total of almost 48,000,000 hops, this approach finds 11 occurrences of patterns of this kind. The number of hops predicted in each of these 11 cases is given in Table~\ref{table2}. 

\begin{table}
\begin{center}
\begin{tabular}{ll}
\textbf{Otter number }& \textbf{Hops predicted} \\
 1 & 267 \\
 2 & 2,235 \\
 3 & 6,840 \\
 4 & 20,463 \\
 5 & 62,895 \\
 6 & 176,040 \\
 7 & 500,271 \\
 8 & 1,387,287 \\
 9 & 3,878,739 \\
10 & 10,830,672 \\
11 & 30,144,672 \\
\end{tabular}
\caption{Observant otter predictions}
\label{table2}
\end{center}
\end{table}

It is clear from this table that the number of hops predicted grows exponentially. 
Note that the final otter step alone predicts 30,144,672 hops, or around 60\% of the total number of hops. 

The general idea is then to search through the execution trace looking for three consecutive matching `shapes', and then checking to see if there is a pattern like the one above, i.e.\ a pattern with (at least) one of the exponents descending. Once we find three occurrences of such a pattern, we then use the three instances to predict the final configuration in this sequence.

This means that we add an extra execution process to the two above for macro machines, which is performed before either of the above steps. This is to search through the execution trace for (at least) two previous configurations that match the current one, and, if found, to calculate the resulting configuration. Otherwise, we proceed as in the two cases above. In some ways, the observant otter may be thought of as a natural counterpart to the process of `decoupling', i.e.\ the second step of the macro machine execution described above. This is because this process essentially changes a string of the form $S^n$ to one of the form $S S^{n-1}$, and hence  decrements $n$. As in the above example for the 5x2 champion, if the process repeated in a predictable pattern, then the observant otter will act to `complete' this process. Put another way, it is this process of `decompressing' $S^n$ to $S S^{n-1}$ (which we shall refer to as the \textit{stretching stork} operation) that produces potential configurations for the observant otter to detect. 

In general, then, if the current configuration is 

$$H_1: X_1^{x_{11}} \ldots X_n^{x_{n1}} \{S\} Y^{y_1} \{D\} Z_1^{z_{11}} \ldots Z_m^{z_{m1}}$$

\noindent
where 
the $X_i$, $Y$ and the $Z_k$ are strings of characters of a fixed length, 
$S$ is the current state,
$D$ the current direction,  and
$H_1$ is the current hop, 
the observant otter searches through the execution trace of the machine for two earlier occurrences of configurations of the same `shape' and for which there is a \textit{regression}, i.e., two configurations of the form 

$$H_2: X_1^{x_{12}} \ldots X_n^{x_{n2}} \{S\} Y^{y_2} \{D\} Z_1^{z_{12}} \ldots Z_m^{z_{m2}}$$

\noindent
and

$$H_3: X_1^{x_{13}} \ldots X_n^{x_{n3}} \{S\} Y^{y_3} \{D\} Z_1^{z_{13}} \ldots Z_m^{z_{m3}}$$

\noindent
where $H_3 < H_2 < H_1$ are the relevant hops and at least one of the following three conditions holds: 

\begin{itemize}
	\item for some $1 \leq u \leq n$ and some $a > 0$ we have $x_{u2} = x_{u1} + a$ and $x_{u3} = x_{u1} + 2a$
	\item for some $a > 0$ we have $y_2 = y_1+a$ and $y_3 = y_1+2a$
	\item for some $1 \leq v \leq m$ and some $a > 0$ we have $z_{v2} = z_{v1} + a$ and $z_{v3} = z_{v1} + 2a$
\end{itemize}

In other words, we have the same sequence of `macro' characters $X_1 \ldots X_n \{S\} Y \{D\} Z_1 \ldots Z_m$ in all three configurations, and the index of at least one of these configurations is regressing as execution proceeds. 

It should be noted that in our implementation we have a slightly stricter criterion that the one above, in that we require either 
 $x_{11} = x_{12} = x_{12}$ or $x_{11} > x_{12} > x_{13}$, i.e.\ we specifically exclude the cases where $x_{11} = x_{12} > x_{13}$ and $x_{11} > x_{12} = x_{13}$. This seems natural, in that in order for a consistent pattern to occur, each element should either remain the same (the first case) or strictly increase each time (the second case). If either of the other cases occur, it is not clear that there is a predictable pattern occurring, and so this case is not of interest for this heuristic. 

In our implementation of the observant otter, we have insisted that exactly one, rather than at least one, of the three conditions hold. This means that there must be exactly one of $X_i$, $Y$ and $Z_j$ for which the index in the second occurrence is less than the corresponding value in the first occurrence. Without loss of generality, we will assume this is $Y$; in this case, the sequence of three configurations will look like this

\begin{center}
\begin{tabular}{ll}
$H_3:$ & $X_1^{x_{13}} \ldots X_n^{x_{n3}} \{S\} Y^{y_1+2a} \{D\} Z_1^{z_{13}} \ldots Z_m^{z_{m3}}$ \\
$H_2:$ & $X_1^{x_{12}} \ldots X_n^{x_{n2}} \{S\} Y^{y_1+a} ~\{D\} Z_1^{z_{12}} \ldots Z_m^{z_{m2}}$ \\
$H_1:$ & $X_1^{x_{11}} \ldots X_n^{x_{n1}} \{S\} Y^{y_1} ~~~~\{D\} Z_1^{z_{11}} \ldots Z_m^{z_{m1}}$ \\
\end{tabular}

\end{center}
where $H_3 < H_2 < H_1$, $x_{i1} \geq x_{i2} \geq x_{i3}$ for all $1 \leq i \leq n$ and $z_{i1} \geq z_{i2} \geq z_{i3}$ for all $1 \leq i \leq m$ 

This `single regressor' assumption simplifies the calculation of the final configuration, at the potential cost of missing some patterns (so that if this property is not satisfied, we do not recognise this pattern and hence keep searching). This is because it is then trivial to calculate how many further occurrences of this pattern occur. Depending on the relative values of $y_1$ and $a$, it is possible that the last element in this sequence has the exponent of the $Y$ term being 0, and in other cases it may not. We calculate the number of further sequences as $y_1 \div a$ unless $y_1$ \textit{mod}$~a = 0$, in which case we use $(y_1 \div a) - 1$.\footnote{Here we use integer division, so $5 \div 2$ is 2.} The reason for this exception is that we choose to always terminate this sequence with the last non-zero exponent for the $Y$ term in the pattern. In the example above, this means that our final configuration is at hop 33120 with the configuration $001(111)^{106}\{b\}(001)^{3}100$ rather than at hop 35097 with the configuration $001{111}^(111)\{b\}(100)$. It may seem more natural to proceed to the zero case when possible, but in some instances this results in incorrect predictions, in that this may cause the heuristic to predict a configuration that does not occur in the `naive' trace.  Hence we have chosen to make the conservative choice of always stopping with the last non-zero exponent for $Y$. Given that this results in comparatively little extra computation, this does not appear to be very costly.  It is an item of ongoing research to determine criteria for when it is `safe' to go to the zero case and when it is not. 



One issue with this approach is that searching through the history of the execution trace has quadratic complexity, and hence has the potential to significantly slow down execution. We address this problem by setting an upper limit on the number of previous steps that are stored. In the results reported below, a `window' of 150 steps was sufficient for all cases tested, including those of the highest productivity. This shows that the patterns detected by the otter are remarkably local, in that it is sufficient to look at most 150 steps in the past to find patterns. Increasing this value may find more patterns, but it comes with a noticeable performance penalty. Whilst we are interested in ensuring that the computation is efficient enough to be feasible, finding a maximally efficient method is outside the scope of this paper. If performance were to become a critical issue, then standard techniques such as hashing or balanced binary trees may be appropriate for improving the performance of this search. 

A further issue is to calculate the number of hops between the current configuration and the predicted one (given that, as above, we have already determined how many occurrences of the pattern found will arise). When the difference between the three occurrences of the pattern is constant, this is straightforward. However, there are cases in which the difference increases, and so some further care is required. If we consider the sequence $H_3$, $H_2$, $H_1$ as the first three terms of an arithmetic sequence $a_0$, $a_1$, $a_2$ ..., then if we can calculate a formula for $a_n$ for all $n \geq 0$, then we only need to compute how many terms of this sequence are appropriate in order to find the predicted number of hops. If the difference is constant, so that $a_2 - a_1 = a_1 - a_0 = d$ for some $d > 0$, then it is simple to see that $a_n = a_0 + n \times d$, and so $a_n = H_3 + n \times d$. In terms of $H_1$, which is our `reference point' for this calculation, we can recast this as $a_m = H_1 + m \times d$, where $m$ is the number of items in the sequence \textbf{after} the occurrence of $H_1$. 

When the difference between the terms is increasing (as in the example above), one way to approach the problem is to find a formula for the sequence of differences, and then use that to find a formula for the hops occurrences. For example, consider again the sequence below. 

\begin{center}
\begin{tabular}{llcc}
& \\
\textbf{Hop} & \textbf{Configuration} & \textbf{Difference} & \textbf{2nd Difference} \\
12393    & $001(111)\{b\}(001)^{66}100$      & --  & --   \\
12480 & $001(111)^{6}\{b\}(001)^{63}100$  & 87  & -- \\
12657    & $001(111)^{11}\{b\}(001)^{60}100$ & 177 & 90 \\
12924    & $001(111)^{16}\{b\}(001)^{57}100$ & 267 & 90 \\
13281 & $001(111)^{21}\{b\}(001)^{54}100$ & 357 & 90 \\
13728    & $001(111)^{26}\{b\}(001)^{51}100$ & 447 & 90 \\
...   & ... & ... & ... \\
27729 & $001(111)^{91}\{b\}(001)^{12}100$ & -- & --\\
29436    & $001(111)^{96}\{b\}(001)^{9}100$  & 1707 & -- \\
31233    & $001(111)^{101}\{b\}(001)^{6}100$ & 1797 & 90\\
33120    & $001(111)^{106}\{b\}(001)^{3}100$ & 1887 & 90 \\
&&&\\
\end{tabular}
\end{center}

The difference between the hops for each otter occurrence is increasing. However, note that the difference between these differences (in the column labelled \textbf{2nd Difference} above) is constant. So if we consider the sequence of differences (the second righthandmost column above), we get $d_1 = 87$ and $d_j = d_1 + (j-1) \times d$ for $j \geq 1$ (where $d$ is the constant second difference). This means that in order to calculate the sequence of hops, we have $a_0 = H_3$, $a_1 = H_2$ and $a_2 = H_1$, as before, but now we have 

$$a_{n+1} = a_n + d_{n+1} = a_n + d_1 + n \times d$$

This is a \textit{linear difference equation} (or \textit{linear recurrence relation}) \cite{Jerri} and its solution is below. 
%
$$a_n = a_0 + d_1 \times n  + d \times n(n-1)/2$$

which we rewrite as 

$$a_n = H_3 + (H_2 - H_3) \times n + (H_1 - H_2) - (H_2 - H_3)) \times n(n-1)/2$$

As above, once the number of occurrences is known, we use this formula to calculate the predicted hop for the final configuration in the otter sequence. As in the previous case, we use $H_1$ as our reference point for this calculation, and so if there are $m > 0$ occurrences of this pattern after $H_1$, the relevant number of hops will be calculated as

$$a_{m+2} = H_3 + (H_2 - H_3) \times (m+2) + (H_1 - H_2) - (H_2 - H_3)) \times (m+2)(m+1)/2$$

In all of the cases we tested, we found that the second differences (i.e.\ the difference between the differences) was always constant, and hence this method will work for all known dreadful dragons. 

\section{Implementation and Results}

We have developed an implementation of both a naive interpreter for Turing machines and one based on macro machines. We then added the observant otter heuristic to the macro machine implementation. This implementation is deliberately simple; there is probably a large amount of scope for improvement, particularly in performance times. In particular, we record the history of execution in a linear list, and hence searching through this list for matching patterns results in a quadratic time search. Clearly this is a rather naive choice, but our aim with this implementation was not to break any `land speed' records, so to speak, but to show that a simple approach together with an appropriate heuristic (in this case, the observant otter) is sufficient to evaluate machines with the largest known productivities to date. 
Our implementation is around 2,000 lines of SWI-Prolog \cite{swi}. Some earlier results from this implementation were reported in \cite{Harland13}; the code has been further developed since then, and a more comprehensive evaluation is reported in this paper. 

Tables~\ref{fig:results1}, \ref{fig:results2}, \ref{fig:results3} and \ref{fig:results4} show all 100 machines that we have evaluated. This list, together with all the code developed, is also available at \url{http://www.cs.rmit.edu.au/~jah/busybeaver}. We have included machines of various dimensions, including 3x3, 6x2, 3x4, 4x3 and 2x6, and one each of dimensions 7x2 and 8x2. 
Many of these machines were taken from Heiner Marxen's web site \cite{Marxen}. Two of the 5-state 2-symbol machines were from Pascal Michel's web site \cite{Michel}. The 7-state and 8-state machines are from Green's 1964 paper \cite{Green64}. 

The results were obtained on a desktop with an Intel i7 with 3.6GHz processors and 8 GB of RAM running Windows 7. This means that these results can be readily reproduced on commodity hardware, and do not require anything more significant than a typical desktop machine. 

In order to investigate the effect of the observant otter, we have also shown in Tables~\ref{fig:results3} and ~\ref{fig:results4} the results of executing the machine without the observant otter heuristic. Naturally for many of the monster machines, it is infeasible to execute the machine to termination using this method, and so there needs to be a bound put on the maximum time taken for such a computation, for which we chose an arbitrary maximum of 5,000,000 steps (i.e.\ 5,000,000 steps in our interpreter, which may correspond to a significantly larger number of hops in the machine execution). This gives us a measure of how much tangible difference the observant otter makes. For some machines, this was sufficient for them to terminate (typically around 75 seconds), and for these we provide no further data. For the other machines (i.e.\ those which require more than 5,000,000 steps to terminate without the otter heuristic), we provide a third point of comparison, which is the result of executing the machine with the otter heuristic but using as a bound the number of hops achieved without the otter heuristic. Specifically, this is the value in the fifth column of Tables~\ref{fig:results3} and ~\ref{fig:results4}. This means that we either have a direct comparison between the computation needed to execute the machine to completion with and without the otter heuristic, or between the computation needed to reach the same stage of computation with and without the otter heuristic. Note that in the latter case, the third computation may have a higher value for the \textbf{Hops} than for the second. This is due to the nature of both macro machines and the observant otter heuristic, which may lead to a computation significantly exceeding the given bound.  However, the computation terminates as soon as it is known that the bound has been exceeded. 

In Tables~\ref{fig:results1}, \ref{fig:results2}, \ref{fig:results3} and \ref{fig:results4} an entry of the form \textbf{(N digits)} denotes a (known) number with \textbf{N} digits. For space reasons, we have adopted the convention of only explicitly writing out numbers in this paper of up to 10 digits in length. Precise values for all results obtained may be found at the above web site. 

~\\
In Tables~\ref{fig:results1} and \ref{fig:results2}: \\
\noindent
The \textbf{No.} entry is our identifier for each machine. \\
The \textbf{Ones} entry is the number of non-blank characters on the tape in the final configuration. \\
The \textbf{Hops} entry is the number of hops needed to reach this configuration (if executed naively). \\
The \textbf{Steps} entry is the number of steps executed in our interpreter. \\
The \textbf{Otters} entry is the number of otter patterns detected during execution. \\
The \textbf{Otter Steps} entry is the number of hops that were predicted by the observant otter heuristic. \\
The \textbf{Otter \%} entry is the percentage of the overall hops that were predicted by the observant otter heuristic. \\
The \textbf{Time} entry is the number of seconds required to reach the final configuration.\\

\begin{table}
\begin{tabular}{lllllllll}
\textbf{No.} & \textbf{Dim.} & \textbf{Ones} & \textbf{Hops} & \textbf{Steps} & \textbf{Otters} & \textbf{Otter Steps} & \textbf{Otter \%} & \textbf{Time}\\
1 & 2x4 & 84 & 6445 & 284 & 2 & 3706 & 57.50 & 0.04 \\
2 & 2x4 & 90 & 7195 & 287 & 2 & 4576 & 63.60 & 0.04 \\
3 & 2x4 & 90 & 7195 & 287 & 2 & 4576 & 63.60 & 0.04 \\
4 & 2x4 & 2050 & 3932964 & 737 & 8 & 3885112 & 98.78 & 0.15 \\
5 & 3x3 & 31 & 2315619 & 824059 & 52 & 420 & 0.02 & 140.28 \\
6 & 3x3 & 5600 & 29403894 & 1050 & 16 & 29254992 & 99.49 & 0.19 \\
7 & 3x3 & 13949 & 92649163 & 467 & 7 & 92440736 & 99.78 & 0.07 \\
8 & 3x3 & 2050 & 3932964 & 737 & 8 & 3885112 & 98.78 & 0.15 \\
9 & 3x3 & 36089 & 310341163 & 859 & 17 & 310094944 & 99.92 & 0.13 \\
10 & 3x3 & 32213 & 544884219 & 739 & 8 & 544398352 & 99.91 & 0.15 \\
11 & 3x3 & 43925 & (10 digits) & 1248 & 14 & (10 digits) & $>$99 & 0.27 \\
12 & 3x3 & 107900 & (10 digits) & 872 & 14 & (10 digits) & $>$99 & 0.20 \\
13 & 3x3 & 43925 & (10 digits) & 1258 & 14 & (10 digits) & $>$99 & 0.27 \\
14 & 3x3 & 1525688 & (12 digits) & 1305 & 29 & (12 digits) & $>$99 & 0.35 \\
15 & 3x3 & 2950149 & (13 digits) & 827 & 13 & (13 digits) & $>$99 & 0.13 \\
16 & 3x3 & 95524079 & (16 digits) & 1149 & 17 & (16 digits) & $>$99 & 0.19 \\
17 & 3x3 & 374676383 & (18 digits) & 4520 & 60 & (18 digits) & $>$99 & 0.70 \\
18 & 5x2 & 4098 & 47176870 & 526 & 11 & 47010381 & 99.65 & 0.12 \\
19 & 5x2 & 4098 & 11798826 & 510 & 10 & 11695761 & 99.13 & 0.07 \\
20 & 5x2 & 4097 & 23554764 & 1456 & 74 & 23480937 & 99.69 & 0.28 \\
21 & 5x2 & 4097 & 11798796 & 498 & 10 & 11721789 & 99.35 & 0.09 \\
22 & 5x2 & 4096 & 11804910 & 550 & 9 & 11686386 & 99.00 & 0.10 \\
23 & 5x2 & 4096 & 11804896 & 544 & 9 & 11686386 & 99.00 & 0.10 \\
24 & 5x2 & 1915 & 2133492 & 393 & 4 & 2076796 & 97.34 & 0.05 \\
25 & 5x2 & 1471 & 2358064 & 931 & 41 & 2312068 & 98.05 & 0.16 \\
26 & 5x2 & 501 & 134467 & 290 & 6 & 121176 & 90.12 & 0.04 \\
27 & 2x5 & 90604 & (10 digits) & 1561 & 9 & (10 digits) & $>$99 & 0.30 \\
28 & 2x5 & 64665 & (10 digits) & 1283 & 10 & (10 digits) & $>$99 & 0.23 \\
29 & 2x5 & 97104 & (10 digits) & 1362 & 17 & (10 digits) & $>$99 & 0.21 \\
30 & 2x5 & 458357 & (12 digits) & 3640 & 84 & (12 digits) & $>$99 & 0.61 \\
31 & 2x5 & 668420 & (12 digits) & 1879 & 12 & (12 digits) & $>$99 & 0.36 \\
32 & 2x5 & 1957771 & (12 digits) & 852 & 12 & (12 digits) & $>$99 & 0.13 \\
33 & 2x5 & 1137477 & (12 digits) & 1559 & 47 & (12 digits) & $>$99 & 0.30 \\
34 & 2x5 & 2576467 & (13 digits) & 1836 & 41 & (13 digits) & $>$99 & 0.45 \\
35 & 2x5 & 4848239 & (14 digits) & 916 & 32 & (14 digits) & $>$99 & 0.25 \\
36 & 2x5 & 143 & 26375397569930 & ?? & ?? & ?? & ?? & ?? \\
37 & 2x5 & 4099 & 15754273 & 958 & 10 & 15654122 & 99.36 & 0.17 \\
38 & 2x5 & 3685 & 16268767 & 765 & 13 & 16171476 & 99.40 & 0.15 \\
39 & 2x5 & 11120 & 148304214 & 947 & 15 & 148035978 & 99.82 & 0.18 \\
40 & 2x5 & 36543045 & (15 digits) & 1685 & 27 & (15 digits) & $>$99 & 0.30 \\
41 & 2x5 & 114668733 & (16 digits) & 961 & 18 & (16 digits) & $>$99 & 0.16 \\
42 & 2x5 & 398005342 & (17 digits) & 648 & 19 & (17 digits) & $>$99 & 0.11 \\
43 & 2x5 & 620906587 & (17 digits) & 919 & 20 & (17 digits) & $>$99 & 0.15 \\
44 & 2x5 & (10 digits) & (18 digits) & 815 & 20 & (18 digits) & $>$99 & 0.15 \\
45 & 2x5 & (10 digits) & (18 digits) & 823 & 20 & (18 digits) & $>$99 & 0.15 \\
46 & 2x5 & (12 digits) & (22 digits) & 862 & 26 & (22 digits) & $>$99 & 0.16 \\
47 & 2x5 & (31 digits) & (62 digits) & 8587 & 168 & (62 digits) & $>$99 & 1.74 \\
48 & 2x5 & (106 digits) & (212 digits) & 24367 & 593 & (212 digits) & $>$99 & 4.54 \\
49 & 2x5 & (106 digits) & (212 digits) & 26169 & 593 & (212 digits) & $>$99 & 4.80 \\
50 & 2x5 & (353 digits) & (705 digits) & 81738 & 1993 & (705 digits) & $>$99 & 15.20 \\
\end{tabular}
\caption{Observant Otter results (part 1)}
\label{fig:results1}
\end{table}
\begin{table}
\begin{tabular}{lllllllll}
\textbf{No.} & \textbf{Dim.} & \textbf{Ones} & \textbf{Hops} & \textbf{Steps} & \textbf{Otters} & \textbf{Otter Steps} & \textbf{Otter \%} & \textbf{Time}\\
51 & 6x2 & 136612 & (11 digits) & 1498 & 15 & (11 digits) & $>$99 & 0.25 \\
52 & 6x2 & 95524079 & (16 digits) & 1514 & 16 & (16 digits) & $>$99 & 0.23 \\
53 & 6x2 & 17485734 & (14 digits) & 1929 & 25 & (14 digits) & $>$99 & 0.30 \\
54 & 6x2 & 36109969 & (16 digits) & 2256 & 83 & (16 digits) & $>$99 & 0.38 \\
55 & 6x2 & 36109970 & (15 digits) & 1000 & 14 & (15 digits) & $>$99 & 0.15 \\
56 & 6x2 & (10 digits) & (20 digits) & 5817 & 215 & (20 digits) & $>$99 & 1.11 \\
57 & 6x2 & (11 digits) & (21 digits) & 1734 & 21 & (21 digits) & $>$99 & 0.26 \\
58 & 6x2 & (14 digits) & (26 digits) & 2155 & 27 & (26 digits) & $>$99 & 0.32 \\
59 & 6x2 & (14 digits) & (28 digits) & 12634 & 408 & (28 digits) & $>$99 & 2.25 \\
60 & 6x2 & (20 digits) & (38 digits) & 4197 & 153 & (38 digits) & $>$99 & 0.73 \\
61 & 6x2 & (22 digits) & (43 digits) & 33228 & 1075 & (43 digits) & $>$99 & 5.73 \\
62 & 6x2 & (48 digits) & (96 digits) & 20493 & 643 & (96 digits) & $>$99 & 3.55 \\
63 & 6x2 & (48 digits) & (96 digits) & 19203 & 641 & (96 digits) & $>$99 & 3.32 \\
64 & 6x2 & (50 digits) & (99 digits) & 28242 & 380 & (99 digits) & $>$99 & 4.61 \\
65 & 6x2 & (61 digits) & (120 digits) & 12321 & 332 & (120 digits) & $>$99 & 1.88 \\
66 & 6x2 & (50 digits) & (100 digits) & 29547 & 389 & (100 digits) & $>$99 & 4.79 \\
67 & 6x2 & (463 digits) & (926 digits) & 265475 & 12330 & (926 digits) & $>$99 & 52.02 \\
68 & 6x2 & (866 digits) & (1731 digits) & 244643 & 14334 & (1731 digits) & $>$99 & 48.22 \\
69 & 6x2 & (882 digits) & (1763 digits) & 288760 & 2211 & (1763 digits) & $>$99 & 44.48 \\
70 & 6x2 & (1440 digits) & (2880 digits) & 406270 & 8167 & (2880 digits) & $>$99 & 65.16 \\
71 & 6x2 & (10567 digits) & (21133 digits) & 1927565 & 99697 & (21133 digits) & $>$99 & 456.05 \\
72 & 6x2 & (18268 digits) & (36535 digits) & 6195809 & 30345 & (36535 digits) & $>$99 & 1154.03 \\
73 & 3x4 & 17323 & 262759288 & 379583 & 17778 & 122527994 & 46.63 & 65.83 \\
74 & 3x4 & (27 digits) & (53 digits) & 9187 & 358 & (53 digits) & $>$99 & 1.98 \\
75 & 3x4 & (141 digits) & (282 digits) & 50694 & 801 & (282 digits) & $>$99 & 9.43 \\
76 & 3x4 & (435 digits) & (869 digits) & 103962 & 3493 & (869 digits) & $>$99 & 21.69 \\
77 & 3x4 & (629 digits) & (1257 digits) & 168263 & 1488 & (1257 digits) & $>$99 & 33.00 \\
78 & 3x4 & (1302 digits) & (2602 digits) & 96341 & 2163 & (2602 digits) & $>$99 & 18.78 \\
79 & 3x4 & (2356 digits) & (4711 digits) & 3178530 & 107045 & (4711 digits) & $>$99 & 559.23 \\
80 & 3x4 & (2373 digits) & (4745 digits) & 944360 & 13465 & (4745 digits) & $>$99 & 156.75 \\
81 & 3x4 & (6519 digits) & (13037 digits) & 731258 & 13667 & (13037 digits) & $>$99 & 140.72 \\
82 & 4x3 & 15008 & 250096775 & 1594 & 27 & 249578940 & 99.79 & 0.37 \\
83 & 4x3 & (714 digits) & (1427 digits) & 375816 & 20051 & (1427 digits) & $>$99 & 68.39 \\
84 & 4x3 & (810 digits) & (1619 digits) & 924926 & 34272 & (1619 digits) & $>$99 & 167.45 \\
85 & 4x3 & (987 digits) & (1974 digits) & 497189 & 7887 & (1974 digits) & $>$99 & 92.18 \\
86 & 4x3 & (3861 digits) & (7722 digits) & 1759075 & 44810 & (7722 digits) & $>$99 & 313.82 \\
87 & 4x3 & (4562 digits) & (9123 digits) & 1511192 & 25943 & (9123 digits) & $>$99 & 295.55 \\
88 & 4x3 & (4932 digits) & (9864 digits) & 1524523 & 57368 & (9864 digits) & $>$99 & 300.73 \\
89 & 4x3 & (6035 digits) & (12069 digits) & 1319358 & 34262 & (12069 digits) & $>$99 & 224.82 \\
90 & 4x3 & (7037 digits) & (14073 digits) & 1695880 & 25255 & (14073 digits) & $>$99 & 300.14 \\
91 & 4x3 & (7037 digits) & (14073 digits) & 1695879 & 25255 & (14073 digits) & $>$99 & 315.79 \\
92 & 2x6 & 10574 & 94842383 & 948 & 11 & 94674070 & 99.82 & 0.22 \\
93 & 2x6 & 10249 & 98364599 & 712 & 13 & 98206675 & 99.84 & 0.18 \\
94 & 2x6 & 15828 & 493600387 & 1263 & 15 & 492945078 & 99.87 & 0.25 \\
95 & 2x6 & (28 digits) & (55 digits) & 5059 & 149 & (55 digits) & $>$99 & 1.01 \\
96 & 2x6 & (822 digits) & (1644 digits) & 139554 & 2733 & (1644 digits) & $>$99 & 31.66 \\
97 & 2x6 & (4932 digits) & (9864 digits) & 1083249 & 49142 & (9864 digits) & $>$99 & 201.36 \\
98 & 2x6 & (4934 digits) & (9867 digits) & 770686 & 16394 & (9867 digits) & $>$99 & 156.34 \\
99 & 7x2 & 22961 & 197700005 & 241 & 8 & 197585144 & 99.94 & 0.07 \\
100 & 8x2 & (45 digits) & (90 digits) & 2670 & 94 & (90 digits) & $>$99 & 1.47 \\
\end{tabular}
\caption{Observant Otter results (part 2)}
\label{fig:results2}
\end{table}

\noindent
In Tables~\ref{fig:results3} and \ref{fig:results4}: \\
\noindent
The \textbf{No.} entry is our identifier for each machine. \\
The \textbf{Hops1} entry is the number of hops needed to reach the final configuration. \\
The \textbf{Time1} entry is the number of seconds required to reach the final configuration using the observant otter heuristic.  \\
The \textbf{Hops2} entry is the number of hops reached without the observant otter heuristic (up to a maximum of 5,000,000 interpreter steps). \\ 
The \textbf{Time2} entry is the number of seconds required for the computation in the previous column. \\
The \textbf{Hops3} entry (if present) is the number of hops reached with the observant otter heuristic using the figure in the \textbf{Hops2} column as a bound. \\
The \textbf{Time3} entry (if present) is the number of seconds required for the computation in the previous column. \\

This means that if we have been able to find the final configuration without the observant otter, we have a direct comparison between this figure and the time taken with the otter heuristic. Otherwise, we have a direct comparison between computation with and without the otter on up to around 75 seconds worth of computation. 

\begin{table}
\begin{tabular}{llllllll}
\textbf{No.} & \textbf{Dim.} & \textbf{Hops1} &  \textbf{Time1} & \textbf{Hops2} & \textbf{Time2} & \textbf{Hops3} & \textbf{Time3}\\
1 & 2x4 & 6445 & 0.04 & 6445 & 0.01 &  -- &  -- \\
2 & 2x4 & 7195 & 0.04 & 7195 & 0.01 &  -- &  -- \\
3 & 2x4 & 7195 & 0.04 & 7195 & 0.01 &  -- &  -- \\
4 & 2x4 & 3932964 & 0.15 & 3932964 & 0.18 &  -- &  -- \\
5 & 3x3 & 2315619 & 140.28 & 2315619 & 13.15 &  -- &  -- \\
6 & 3x3 & 29403894 & 0.19 & 29403894 & 0.71 &  -- &  -- \\
7 & 3x3 & 92649163 & 0.07 & 92649163 & 0.63 &  -- &  -- \\
8 & 3x3 & 3932964 & 0.15 & 3932964 & 0.18 &  -- &  -- \\
9 & 3x3 & 310341163 & 0.13 & 310341163 & 3.03 &  -- &  -- \\
10 & 3x3 & 544884219 & 0.15 & 544884219 & 1.58 &  -- &  -- \\
11 & 3x3 & (10 digits) & 0.27 & (10 digits) & 4.86 &  -- &  -- \\
12 & 3x3 & (10 digits) & 0.20 & (10 digits) & 5.09 &  -- &  -- \\
13 & 3x3 & (10 digits) & 0.27 & (10 digits) & 5.09 &  -- &  -- \\
14 & 3x3 & (12 digits) & 0.35 & (12 digits) & 74.85 & (12 digits) & 0.28\\
15 & 3x3 & (13 digits) & 0.13 & (12 digits) & 73.87 & (13 digits) & 0.13\\
16 & 3x3 & (16 digits) & 0.19 & (12 digits) & 73.37 & (13 digits) & 0.14\\
17 & 3x3 & (18 digits) & 0.70 & (12 digits) & 77.25 & (12 digits) & 0.44\\
18 & 5x2 & 47176870 & 0.12 & 47176870 & 0.40 &  -- &  -- \\
19 & 5x2 & 11798826 & 0.07 & 11798826 & 0.21 &  -- &  -- \\
20 & 5x2 & 23554764 & 0.28 & 19988787 & 77.40 & 23468894 & 0.27\\
21 & 5x2 & 11798796 & 0.09 & 11798796 & 0.22 &  -- &  -- \\
22 & 5x2 & 11804910 & 0.10 & 11804910 & 0.21 &  -- &  -- \\
23 & 5x2 & 11804896 & 0.10 & 11804896 & 0.21 &  -- &  -- \\
24 & 5x2 & 2133492 & 0.05 & 2133492 & 0.07 &  -- &  -- \\
25 & 5x2 & 2358064 & 0.16 & 2358064 & 6.15 &  -- &  -- \\
26 & 5x2 & 134467 & 0.04 & 134467 & 0.03 &  -- &  -- \\
27 & 2x5 & (10 digits) & 0.30 & (10 digits) & 8.03 &  -- &  -- \\
28 & 2x5 & (10 digits) & 0.23 & (10 digits) & 7.56 &  -- &  -- \\
29 & 2x5 & (10 digits) & 0.21 & (10 digits) & 12.14 &  -- &  -- \\
30 & 2x5 & (12 digits) & 0.61 & (11 digits) & 77.29 & (11 digits) & 0.58\\
31 & 2x5 & (12 digits) & 0.36 & (12 digits) & 59.31 &  -- &  -- \\
32 & 2x5 & (12 digits) & 0.13 & (12 digits) & 74.43 & (12 digits) & 0.13\\
33 & 2x5 & (12 digits) & 0.30 & 5999830 & 76.38 & 15516871 & 0.14\\
34 & 2x5 & (13 digits) & 0.45 & (12 digits) & 73.92 & (12 digits) & 0.41\\
35 & 2x5 & (14 digits) & 0.25 & (12 digits) & 75.72 & (12 digits) & 0.21\\
36 & 2x5 & ?? & ?? & ?? & ?? & ?? & ??\\
37 & 2x5 & 15754273 & 0.17 & 15754273 & 0.43 &  -- &  -- \\
38 & 2x5 & 16268767 & 0.15 & 16268767 & 0.46 &  -- &  -- \\
39 & 2x5 & 148304214 & 0.18 & 148304214 & 1.37 &  -- &  -- \\
40 & 2x5 & (15 digits) & 0.30 & (12 digits) & 78.99 & (12 digits) & 0.22\\
41 & 2x5 & (16 digits) & 0.16 & (12 digits) & 73.88 & (12 digits) & 0.11\\
42 & 2x5 & (17 digits) & 0.11 & (12 digits) & 71.27 & (12 digits) & 0.07\\
43 & 2x5 & (17 digits) & 0.15 & (12 digits) & 72.85 & (13 digits) & 0.11\\
44 & 2x5 & (18 digits) & 0.15 & (12 digits) & 72.60 & (12 digits) & 0.09\\
45 & 2x5 & (18 digits) & 0.15 & (12 digits) & 72.11 & (12 digits) & 0.09\\
46 & 2x5 & (22 digits) & 0.16 & (12 digits) & 72.71 & (13 digits) & 0.08\\
47 & 2x5 & (62 digits) & 1.74 & (12 digits) & 76.80 & (12 digits) & 0.25\\
48 & 2x5 & (212 digits) & 4.54 & (12 digits) & 75.46 & (12 digits) & 0.21\\
49 & 2x5 & (212 digits) & 4.80 & (12 digits) & 75.95 & (12 digits) & 0.23\\
50 & 2x5 & (705 digits) & 15.20 & (12 digits) & 78.16 & (12 digits) & 0.20\\
\end{tabular}
\caption{Further Observant Otter results (part 1)}
\label{fig:results3}
\end{table}
\begin{table}
\begin{tabular}{llllllll}
\textbf{No.} & \textbf{Dim.} & \textbf{Hops1} &  \textbf{Time1} & \textbf{Hops2} & \textbf{Time2} & \textbf{Hops3} & \textbf{Time3}\\
51 & 6x2 & (11 digits) & 0.25 & (11 digits) & 7.86 &  -- &  -- \\
52 & 6x2 & (16 digits) & 0.23 & (12 digits) & 78.61 & (13 digits) & 0.17\\
53 & 6x2 & (14 digits) & 0.30 & (12 digits) & 80.22 & (13 digits) & 0.25\\
54 & 6x2 & (16 digits) & 0.38 & 39940703 & 76.13 & 173556244 & 0.17\\
55 & 6x2 & (15 digits) & 0.15 & (13 digits) & 77.64 & (14 digits) & 0.13\\
56 & 6x2 & (20 digits) & 1.11 & 19943326 & 77.26 & 29765228 & 0.26\\
57 & 6x2 & (21 digits) & 0.26 & (13 digits) & 77.50 & (14 digits) & 0.14\\
58 & 6x2 & (26 digits) & 0.32 & (13 digits) & 76.99 & (14 digits) & 0.14\\
59 & 6x2 & (28 digits) & 2.25 & 39848493 & 79.99 & 57854154 & 0.42\\
60 & 6x2 & (38 digits) & 0.73 & 54760346 & 76.58 & 252160142 & 0.11\\
61 & 6x2 & (43 digits) & 5.73 & 39839230 & 81.07 & 43619704 & 0.57\\
62 & 6x2 & (96 digits) & 3.55 & 48524764 & 77.41 & 92937448 & 0.20\\
63 & 6x2 & (96 digits) & 3.32 & 32668321 & 82.85 & 41281924 & 0.17\\
64 & 6x2 & (99 digits) & 4.61 & (12 digits) & 79.83 & (12 digits) & 0.42\\
65 & 6x2 & (120 digits) & 1.88 & (13 digits) & 76.50 & (13 digits) & 0.16\\
66 & 6x2 & (100 digits) & 4.79 & (12 digits) & 77.90 & (12 digits) & 0.42\\
67 & 6x2 & (926 digits) & 52.02 & 19932648 & 73.94 & 25237532 & 0.54\\
68 & 6x2 & (1731 digits) & 48.22 & 22225873 & 75.71 & 39845896 & 0.14\\
69 & 6x2 & (1763 digits) & 44.48 & (13 digits) & 77.09 & (13 digits) & 0.23\\
70 & 6x2 & (2880 digits) & 65.16 & (12 digits) & 80.84 & (12 digits) & 0.20\\
71 & 6x2 & (21133 digits) & 456.05 & 19959284 & 76.63 & 36213129 & 0.13\\
72 & 6x2 & (36535 digits) & 1154.03 & (12 digits) & 79.13 & (13 digits) & 0.48\\
73 & 3x4 & 262759288 & 65.83 & 50789361 & 98.93 & 50798489 & 28.62\\
74 & 3x4 & (53 digits) & 1.98 & 6005442 & 73.57 & 11408546 & 0.18\\
75 & 3x4 & (282 digits) & 9.43 & (12 digits) & 76.52 & (12 digits) & 0.43\\
76 & 3x4 & (869 digits) & 21.69 & 19983565 & 80.35 & 55369106 & 0.12\\
77 & 3x4 & (1257 digits) & 33.00 & (12 digits) & 76.21 & (12 digits) & 0.32\\
78 & 3x4 & (2602 digits) & 18.78 & (13 digits) & 71.21 & (14 digits) & 0.10\\
79 & 3x4 & (4711 digits) & 559.23 & 11974138 & 75.92 & 12292474 & 0.46\\
80 & 3x4 & (4745 digits) & 156.75 & (12 digits) & 74.23 & (12 digits) & 0.25\\
81 & 3x4 & (13037 digits) & 140.72 & (12 digits) & 71.51 & (13 digits) & 0.17\\
82 & 4x3 & 250096775 & 0.37 & 250096775 & 1.94 &  -- &  -- \\
83 & 4x3 & (1427 digits) & 68.39 & 13319120 & 72.58 & 17502670 & 0.21\\
84 & 4x3 & (1619 digits) & 167.45 & 6264241 & 74.35 & 8863780 & 0.45\\
85 & 4x3 & (1974 digits) & 92.18 & (12 digits) & 72.28 & (12 digits) & 0.40\\
86 & 4x3 & (7722 digits) & 313.82 & 15567444 & 75.15 & 24601805 & 0.20\\
87 & 4x3 & (9123 digits) & 295.55 & (12 digits) & 77.07 & (12 digits) & 0.26\\
88 & 4x3 & (9864 digits) & 300.73 & 19934942 & 81.68 & 37620302 & 0.16\\
89 & 4x3 & (12069 digits) & 224.82 & (12 digits) & 75.49 & (12 digits) & 0.16\\
90 & 4x3 & (14073 digits) & 300.14 & (12 digits) & 74.43 & (13 digits) & 0.21\\
91 & 4x3 & (14073 digits) & 315.79 & (12 digits) & 73.11 & (13 digits) & 0.21\\
92 & 2x6 & 94842383 & 0.22 & 94842383 & 1.23 &  -- &  -- \\
93 & 2x6 & 98364599 & 0.18 & 98364599 & 1.04 &  -- &  -- \\
94 & 2x6 & 493600387 & 0.25 & 493600387 & 2.83 &  -- &  -- \\
95 & 2x6 & (55 digits) & 1.01 & (12 digits) & 74.71 & (12 digits) & 0.18\\
96 & 2x6 & (1644 digits) & 31.66 & (12 digits) & 77.38 & (13 digits) & 0.26\\
97 & 2x6 & (9864 digits) & 201.36 & 13328142 & 77.74 & 48673870 & 0.11\\
98 & 2x6 & (9867 digits) & 156.34 & (12 digits) & 72.78 & (13 digits) & 0.25\\
99 & 7x2 & 197700005 & 0.07 & 197700005 & 1.31 &  -- &  -- \\
100 & 8x2 & (90 digits) & 1.47 & (12 digits) & 71.64 & (13 digits) & 0.14\\
\end{tabular}

\caption{Further Observant Otter results (part 2)}
\label{fig:results4}
\end{table}

We re-iterate that developing the fastest possible interpreter for such machines is not our primary aim. Clearly efficiency is an important factor, and the measurement of execution time is vital for comparative purposes. It is also important to keep in mind that the endless nature of the busy beaver problem (i.e.\ that whatever the current state of knowledge, there is always a larger problem instance to be addressed) means that efficiency is both fundamentally important, as greater efficiency implies greater exploration of the problem,  and ultimately futile, in that there will always be a larger problem size just beyond the reach of the current state of technology. 
However, the most important point to note for the purposes of this paper is that we can execute all but one of these machines (with the exception of machine 36, see below) in feasible times on commodity hardware. The longest execution time was around 20 minutes (machine 72), with all but 12 machines (numbers 5, 36, 71, 72, 79, 86-91, 97) taking less than 200 seconds each. Other than machine 36, the entire set of machines can be run to termination within 90 minutes, and our entire suite of tests in just under three hours. Whilst these times can certainly be improved, the most important aspect is that the use of the observant otter heuristic can make seemingly vast computations feasible. It should also be noted that the observant otter was able to predict over 97\% of the computation steps for all but 7 machines. 

\section{Discussion}\label{sec:discussion}

As can be seen from Tables~\ref{fig:results1} and~\ref{fig:results2}, for the machines of very high productivity, the observant otter has proved to be very effective, with the observant otter usually predicting at least 99\% of the hops required for termination. It also shows some potentially surprising results, in that the machines with the largest productivities weren't necessarily the ones which required the largest number of observant otter occurrences. For example, machine 18, the 5x2 champion, required 11 occurrences of the observant otter, but machine 20 required 74. This suggests that the number of such occurrences is a more intuitive measure of the difficulty of execution than the sheer size of the final configuration or the number of hops required to compute it. 

It is also worth noting that the ratio of \textbf{Hops} to \textbf{Steps} in the above table is very high, especially for the machines of largest productivity. Clearly the \textbf{Steps} value increases as the productivity does, but at an exponentially lower rate than \textbf{Hops}. 

A key decision to make in the implementation is to determine how large the `history window' should be. After a little experimentation, it was found that 150 works for all the above cases, but making it smaller tended to make the performance deteriorate significantly on some of the larger machines. This shows that the otter patterns are remarkably local, in that despite the large number of steps involved, only a small fraction of the most recent ones are needed to detect patterns (or at least, the patterns detected in this way are sufficient to feasibly execute the machine). As mentioned above, it is possible to improve the efficiency of this procedure by using standard data structures such as hash tables or balanced binary trees rather than the simple list implementation used here. 

It is also important to note that whilst our implementation only applied the observant otter heuristic when there was only a single regressor, it did record when multiple regressors were noticed (and hence provide some scope for further optimisation by exploiting this property). Of the 100 machines tested, multiple regressors were detected in only four machines (numbers 36, 56, 84 and 93). This means that a more sophisticated approach to calculating the final configuration for the observant otter (by exploiting the multiple regressor property rather than ignoring it) may increase its effectiveness for at most four machines out of 100. In addition, two of these four (numbers 56 and 93) take only 1.12 and 0.19 seconds to execute at present, which significantly limits the scope for improvements in execution time. Clearly, though, reducing the execution time for machine 84 (presently 177.57 seconds) via this method may be useful. 

A further potential improvement to this process is to store the otter patterns as they are found, rather than computing them `afresh' each time. For example, the 11 otter occurrences in machine 18 are basically different instances of two particular otter patterns. However, storing and applying the correct pattern is not altogether straightforward, particularly as the number of steps in each pattern needs to be calculated differently. This is an item of future work. 

A related aspect is that of finding the `earliest possible' otter pattern. To return to the 5x2 champion yet again, the first otter detected by our process is the one below. 

\begin{center}
\begin{tabular}{ll}
& \\
\textbf{Hop} & \textbf{Configuration }\\
448 & $001111\{b\}(001)^{12}\{l\}100$ \\
535 & $001(111)^{6}\{b\}(001)^{9}\{l\}100$ \\
712 & $001(111)^{11}\{b\}(001)^{6}\{l\}100$ 	 \\ 
& \\
\end{tabular}
\end{center}

However, there are some earlier patterns that the macro machine structure makes it difficult to detect. This is because the choice of $k=3$ is generally best for this machine (due to strings such as $(001)^{522}$ occurring in the execution trace), but this has the effect of representing a string of 18 1's as $(111)^{6}$, rather than the more compact and intuitive form $1^{18}$. In order to find the earliest possible occurrences of patterns, it thus seems necessary to develop an adaptive compression of configurations, so that we can represent the above sequence of configurations as below.  

\begin{center}
\begin{tabular}{ll}
& \\
\textbf{Hop} & \textbf{Configuration }\\
448 & $1^{4} \{b\}(001)^{12}\{l\}1$ \\
535 & $1^{19}\{b\}(001)^{9}\{l\}1$ \\
712 & $1^{34}\{b\}(001)^{6}\{l\}1$ \\ 
& \\
\end{tabular}
\end{center}

This finer granularity will enable `earlier' otter occurrences to be detected. This is also an item of future work. 

It also seems worth remarking that often the size of the otter steps predicted seems to increase exponentially as the computation goes on. We saw above how the final otter step alone for the 5x2 champion predicts 30,144,672 steps, or around 60\% of the total number of steps. Clearly there is an exponential growth in the number of predicted steps, yet the overall computation still terminates. This property of terminating growth is the inspiration for the name \textit{leashed leviathans}; there is an exponential computation happening here, but it reaches a limit at some point. This behaviour is reminiscent of the famous $3n+1$ sequence \cite{Michel93,Michel13}. For now, we note that the observant otter may be thought of as a means of coping with this behaviour. 

It is also worth noting that for the machine of very high productivity, the number of \textbf{Hops} taken seems to be almost always the square of the number of non-blanks in the final configuration, and that this property seems remarkably consistent, despite the very large numbers involved \cite{LP07}. There are more sophisticated analyses of busy beaver champions than we have given here (such as those by Michel \cite{Michel}); here we remark that it seems intriguing to contemplate the connection between the patterns found by the observant otter and this property, particularly as the machines that adhere most to this property tend to be ones for which the observant otter is particularly effective.  

As mentioned above, in the above results we always applied the otter in such a way that the predicted configuration was always one less than the maximum possible. In other words, given a pattern such as $X^n Y^m$ leading to $X^{n+2} Y^{m-1}$, our calculation always predicted the final state as $X^{n+2(m-1)} Y$ rather than the potentially more efficient choice of $X^{n+2m}$. Whilst this approach worked in many cases, there are some in which this approach leads to erroneous results. Determining precise criteria under which one can `safely' use the zero case is also an item of future work. This seems particularly important for the patterns occurring in a non-termination context; when used on terminating machines, our practice is technically less efficient, but only results in a very small increase in the time taken to execute the machine. 


It should also be noted that the observant otter is not a panacea; in fact there were two machines where the otter was of little or no use (machines 5 and 36), and one where its use was significantly less significant than many others (machine 73). These three machines were three of only seven in which the number of hops predicted by the observant otter was less than 98\% of the total; of the remaining four, three (machines 1, 2 and 3) were of sufficiently small computation size that the observant otter executed more slowly then the naive version, and for the fourth (machine 26), the observant otter `only' predicted 90\% of the hops. 
Machines 5 and 36 of these are remarkable for their non-adherence to the above mentioned `square law', in that these machines have productivity 31 and activity 2,315,619, and productivity 143 and activity 26,375,397,569,930 respectively. These machines do not produce configurations with large exponents (e.g.\ containing strings like $(001)^{123194244242}$), and in fact do not have any patterns recognised by the observant otter. 
These two machines are examples of what we call {\em wandering wastrels}; these are machines that have very low productivities compared to activity  (i.e.\ the productivity is very much lower than the square root of the activity). Machine 73 is a hybrid case, in that almost half the steps were predicted by the observant otter, but the other half were not.
No doubt there are other varieties of extreme machine behaviour `out there', so that when dealing with a systematic analysis of a large number of machines, it will be necessary to have a number of different techniques in addition to the observant otter. However, we believe that the results of this paper have shown that the observant otter is a crucial component of the techniques required. 

It is a matter of some regret that we were unable to verify that machine 36 in fact terminates at all; all our attempts to execute it to termination failed. Hence stated productivity and hops required to reach the termination state are taken from Marxen's web page, but we have been unable to verify these. The main issue is that the strings produced during the execution of this machine are not compressible in the same way that most others are. For example, consider the part of the execution trace below. 

\begin{center}
\begin{tabular}{ll}
\textbf{Hop} & \textbf{Configuration} \\
& \\
$10452$ & $011110101111 (10)^5 (11)  \{b\}(14) 34334343$ \\
$10468$ & $011110101111 (10)^4 (11)^2\{b\}(14) 44334343$ \\
$10510$ & $011110101111 (10)^3 (11)^3\{b\}(14) 43434343$ \\   
\end{tabular}
\end{center}
 
Hops 10452 and 1068 suggest that there may be an otter pattern here. However, hop 10510 does not fit this pattern, as the string to the right of $14$ does not have an identifiable pattern, unlike the string to the left of it. A little later in the trace we find a similar sequence below. 

\begin{center}
\begin{tabular}{ll}
\textbf{Hop} & \textbf{Configuration} \\
&\\
$10830$ & $0111(10)^2 (11)^3 (10)^4 (11)  \{b\}(14)343 33443$ \\
$10846$ & $0111(10)^2 (11)^3 (10)^3 (11)^2\{b\}(14)443 33443$ \\
$10888$ & $0111(10)^2 (11)^3 (10)^2 (11)^3\{b\}(14)434 33443$ \\ 
\end{tabular}
\end{center}

Again, the first two hops suggest that there may be an otter pattern, but the third does not fit this. The main issue seems to be that the string to the right of $14$ gets longer throughout the computation, but does so in a way that is not easily compressed, and hence it is not recognised by the observant otter heuristic. 
This suggests that whilst the observant otter works well on highly compressible patterns, such as $(110)^{34}$, there is a need for a different technique as well in order to deal with strings which are more random and hence less compressible than those typically found in the dreadful dragons in this paper. 

One point that should be noted is that the observant otter heuristic is also crucial to the success of non-termination proofs. As discussed in \cite{Harland07}, a key aspect of many non-termination proofs is to determine a pattern from the execution trace, and then show that this pattern will recur infinitely often. This is usually done by showing that a pattern is growing in size, such as observing the sequence of configurations below, 

$$11\{c\}11$$ 
$$(11)^2\{c\}11$$
$$(11)^3\{c\}11$$

and inferring from them that a configuration of the form $(11)^{n}\{c\}11$ will eventually result in one of the form $(11)^{n+1}\{c\}11$ for any $n \geq 1$. This process is certainly similar to the observant otter, but seems a little distinct from it (not the least because the count is always increasing here). However, the observant otter patterns often crop up when attempting to show that configurations such as $(11)^{n}\{c\}11$ eventually result in the configuration $(11)^{n+1}\{c\}11$. 

\begin{figure}
	\centering
				\includegraphics[width=15cm]{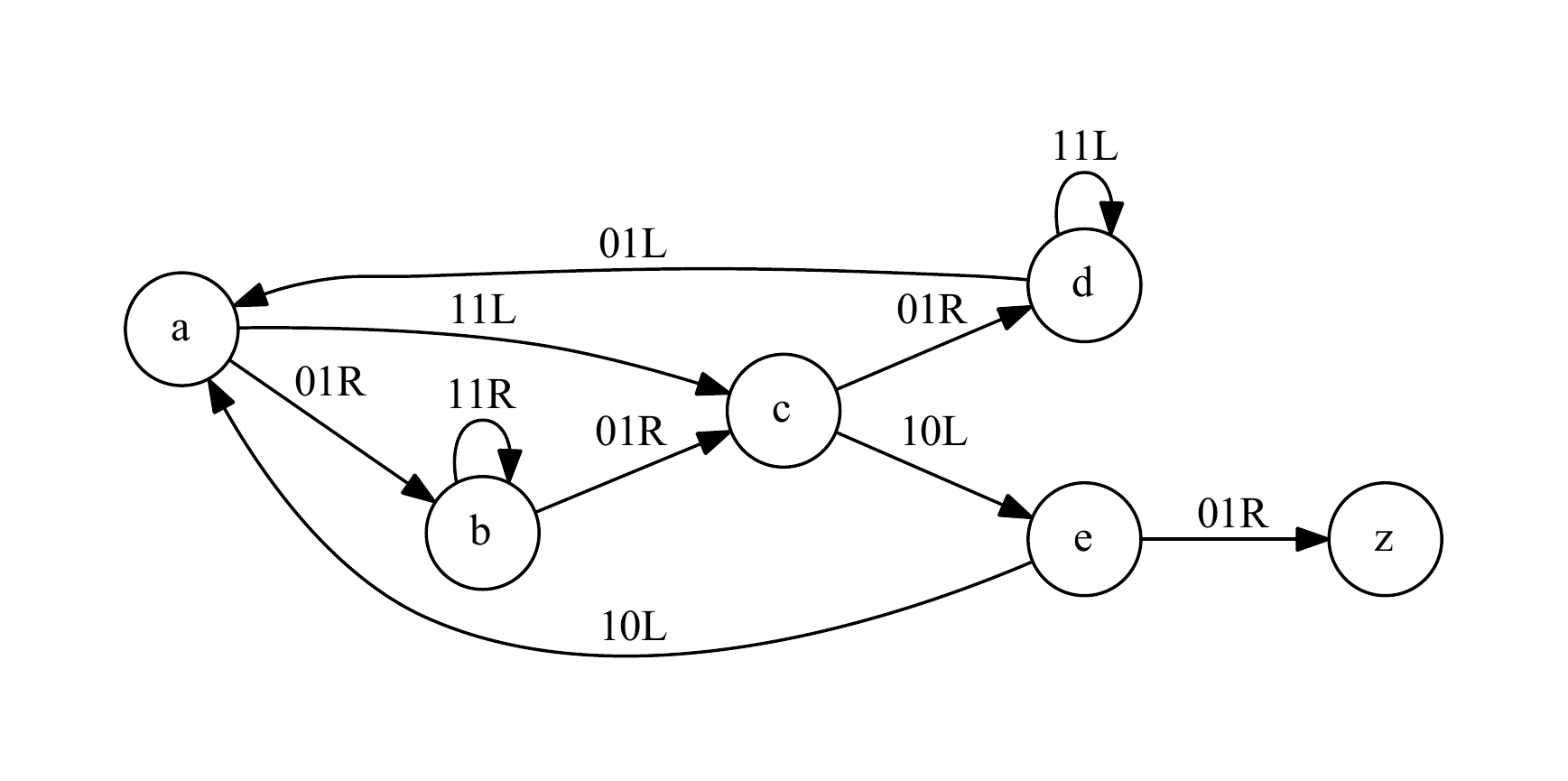}
			\caption{Non-terminating machine}
	\label{fig:example3}
\end{figure}

For example, consider the machine in Figure~\ref{fig:example3}.

The trace of the execution of this machine contains the following configurations.  

$$\{c\}11$$     
$$11\{c\}11$$     
$$   10\{c\}111$$    
$$   1111\{c\}11$$    
$$  1011\{c\}111 $$    
$$  110\{c\}1111  $$   
$$  111111\{c\}11 $$
$$ 101111\{c\}111  $$   
$$ 11011\{c\}1111  $$   
$$ 1110\{c\}11111  $$   
$$ 11111111\{c\}11  $$   

From this it is straightforward to see that patterns of the form $(11)^m \{c\}11$ will continue indefinitely (there are four instances in the segment shown above).  Hence we attempt to show that the configuration $(11)^{n}\{c\}1$ (for an arbitrary $n$) will eventually lead to $(11)^{n+1}\{c\}1$. In doing so, a necessary intermediate step is to go from the configuration $1^n 0 1^k \{c\} 1^m$ to $1^{n+1} 0 1^{k-2} \{c\} 1^{m+1}$, which is precisely an observant otter step. Hence, we need the observant otter to determine that the machine will eventually reach the configuration $1^{n+l} 0 1^{j} \{c\} 1^{m+l}$ where $l = k \div 2$ and $j = k \; mod \; 2$, and from there to $(11)^{n+2}\{c\}1$, at which point we can conclude that we have successfully shown that this machine does not terminate. 
Hence the observant otter seems to be fundamental to the execution of these machines, whether terminating or not. 

A further observation is that in principle, it may be possible to infer the otter patterns from an analysis of the machine, rather than by observing execution. This would presumably be significantly more efficient than the current quadratic search that is used in our implementation. However, it is likely that this would be possible for a particular class of machines, meaning that it will probably always be necessary to fall back on a similar (but more efficient) method like the one described here. 


One curiosity of note is that all of the above dreadful dragons except machine 92 are \textit{exhaustive} machines, i.e.\ that if the machine dimension is $d$, then there is exactly one halting transition and $d-1$ standard transitions. Machine 92 has two halt transitions (one of which will never be used by computation on the blank input), but the point to note is that it is the only non-exhaustive machine amongst the ones we have found. Given that it is not of particularly large productivity for a 2-state 6-symbol machine, this may simply reflect the lack of exploration of the very large space of machines of dimension 12. 

As mentioned above, once the busy beaver problem is settled for a certain dimension, we may use the output produced to address the placid platypus problem, which is to determine the machine with the smallest number of states that will produce an output of a given length. For this to have a non-trivial answer, we need to specify the number of symbols used in the machine in advance. A well-known result of Shannon has shown that any Turing machine can be transformed into an equivalent one with exactly two states \cite{Shannon} (but with a significantly larger number of symbols), so that unless the number of symbols is restricted, the answer will always be 2. Hence for a given number of symbols $m$, the placid platypus problem for $k$ is to find the smallest $n$ such that an $n$-state $m$-symbol Turing machine terminates on the blank input with $k$ non-blank characters on the tape. We will use the notation $pp(k,m)$ to mean the placid platypus value for $k$ non-blank characters and $m$ symbols.  It is obvious that $p(k,m) \leq k$ for any $m \geq 2$.\footnote{We call an $n$-state machine of productivity $n$ a \textit{gutless goanna}.}
For the 2-symbol case, as $bb(2,2) = 4$, $bb(3,2) = 6$ and $bb(4,2) = 13$, we know that $pp(4,2) = 2$, $pp(6,2) = 3$, $pp(k,2) \geq 4$ for $k \geq 7$, and $pp(k,2) \geq 5$ for $k \geq 14$. Inspection of the class of machines shows that there is a 2-state 2-symbol machine of productivity 3, and a 3-state 2-symbol machine of productivity 5, and that there is a 4-state 2-symbol machine for every productivity between 1 and 13. This is summarised in the table below.

\begin{center}
\begin{tabular}{lllllllllllllll}
&&&&&&&&&&&&&\\
$k$ & 1 & 2 & 3 & 4 & 5 & 6 & 7 & 8 & 9 & 10 & 11 & 12 & 13 & $\geq$ 14 \\
$pp(k,2)$ & 1 & 2 & 2 & 2 & 3 & 3 & 4 & 4 & 4 & 4 & 4 & 4 & 4 & $\geq 5$ \\
\end{tabular}
\end{center}

The case for $pp(5,2)$ is more interesting, in that it seems very likely that $bb(5,2) = 4098$, and we also know that there are 5-state 2-symbol machines of productivity 4097, 4096, 1915, 1471 and 501, but seemingly no other ones of productivity above 500. This raises the interesting question of the smallest number which is not the productivity of a 5-state 2-symbol machine, i.e.\ the smallest $k$ such that $pp(k,2) \geq 6$. A full answer to this and similar questions will only be possible once this class of machines has been fully explored. 



\section{Conclusions and Further Work}

We have seen how the busy beaver problem is simple and elegant to state, but that solving it, even for small dimensions, is a much more complex task. We have seen that whilst there is a large amount of knowledge about this problem that there is a need for a rigourous approach to the provision of evidence for claims made, and for a methodological process that invites scrutiny and facilitates collaboration. The provision of this evidence will also allow a number of other properties of interest to be evaluated. 
We have seen how the observant otter heuristic, even when implemented in a simple manner, makes it possible to efficiently evaluate machines of very high productivity. We have also seen how this heuristic seems to be a natural extension of the implementation of macro machines, and is also appropriate for evaluating machines in general, terminating or otherwise. 
It is our intention to develop methods to execute and analyse machines, and then use these methods to address the busy beaver and related problems. Hence we see this paper as a first step in a longer process to develop analysis and generation methods which will then be applied to all the categories above, up to and including the Demon Duck of Doom. However, there is still a significant amount of work to be done to achieve this.  

As noted above, our implementation can be improved in various ways. One obvious such improvement is to use a more efficient data structure to store the execution history, so that the search time for matching patterns can be significantly reduced. 

It still remains an open issue as to how we may determine the earliest possible application of the observant otter. As mentioned above, this is related to the design of macro machines, and to solve this problem will require a more flexible machine architecture. Rather than pick some values in advance, and see which ones work best (which is the current approach), one improvement would be to run a (naive) execution for say 1,000 steps, examine the trace for patterns, and to determine the most appropriate representation. An alternative approach is to abandon the use of a fixed block size, and to dynamically adapt the representation as the machine executes. This adaptive compression approach may also be helpful for wandering wastrels such as machine 36. This is an item of future work. 

Further work is also needed in order to determine precise conditions under which the predicted configuration can be safely calculated with a value of 0 for the final value of the regressor.
As mentioned above, for terminating machines, there is not a great loss of efficiency in being conservative and always avoiding this case. For non-terminating machines, it may be critical to be more precise. 

Another avenue of further work is to be able to store patterns, and hence be able to reuse them without having to `rediscover' patterns. We refer to this process as the {\em ossified ocelot} heuristic, which will be the subject of a future paper. Whilst it seems an intuitively natural thing to do, a key problem is to know exactly what to store, i.e.\ which parts of a given configuration are strictly necessary to the pattern, and which are irrelevant. It seems also that the number of `different' otter patterns necessary for the evaluation of the machine is an even more intuitive measure of the complexity of the machine than the number of otter applications used during evaluation. 

Another line of future work is to generalise the observant otter to incorporate multiple regressors. This seems intuitively simple; when such an occurrence is detected, determine which of the decreasing counters will first run out, and then proceed as before, with the only difference being that some other values will decrease rather than increase. It remains to be seen how much difference this will make in practice. This also may complicate the analysis of when it is safe to use a final value of 0 for the (minimal) regressor. 

Another item of note is that in all the cases we found, the process for determining the number hops predicted by the otter always found that the second differences were constant. This suggests that there may be an analytical result that can be proved along these lines, such as that for any machine of up to a given dimension, the second differences between otter patterns is always constant, and hence the hops can be predicted by the above quadratic formula. It seems intuitive that for some sufficiently large machine that the second differences will no longer be constant (due to more complex patterns emerging), and that some larger formula (presumably cubic) will be needed. However, it is not clear that this point has been reached yet. 

As noted earlier, efficiency is not one of our primary concerns. However, it cannot be totally ignored, and given the endless nature of the busy beaver problem, continuous refinement and improvement of the techniques used will be necessary. It should be kept in mind that the dreadful dragons evaluated in this paper are ones already known, and hence are like large animals bred in captivity, in that they are known to terminate and have often been independently evaluated by other researchers. Given the tendency for 2-state $n$-symbol machines to be more complex than $n$-state 2-symbol ones, and that the current record machines have 6 states and 2 symbols, it seems that there are machines of yet larger productivities `out in the wild', i.e.\ the unexplored spaces of the machines of dimension 12.  
Who knows what fearsome monsters (\textit{baleful balrogs} or \textit{devious demons}) lurk just beyond the edges of the known map? 

\section*{Acknowledgements}

The author would like to thank
Allen Brady, 
Jeanette Holkner, 
Barry Jay, 
Andy Kitchen, 
Pascal Michel,
Sandra Uitenbogerd, 
Michael Winikoff,
and some anonymous referees for valuable feedback on this material. 

\section*{References}

\bibliographystyle{elsarticle-num} 
\bibliography{busybeaver}

\end{document}